%% file: ms_xar.tex
\documentclass[manuscript]{aastex}
\usepackage{emulateapj5, apjfonts, epsfig}

\def\ros{{\sl ROSAT }}

\def\ein{{\sl Einstein }}

\def\chandra{{\sl Chandra }}

\def\ergsec{\hbox{erg s$^{-1}$ }}
\def\ergcm{\hbox{erg cm$^{-2}$ s$^{-1}$ }}

\def\Msun{$M_{\odot}$ }
\def\Rsun{$R_{\odot}$ }
\def\Lsun{$L_{\odot}$ }

\def\nvii{N~{\sc vii}}
\def\ovii{O~{\sc vii}}
\def\oviii{O~{\sc viii}}

\def\fexviii{Fe~{\sc xviii}}
\def\neix{Ne~{\sc ix}}

\def\mgxi{Mg~{\sc xi}}

\slugcomment{submitted for publication in The Astrophysical Main Journal}
\shorttitle{PMS-STARS IN THE ORION TRAPEZIUM}
\shortauthors{SCHULZ et al.}

\begin{document}

\title{X-ray Properties of Low-Mass Pre-main Sequence Stars in the Orion Trapezium Cluster}

\author{
Norbert S. Schulz\altaffilmark{1}, 
David P. Huenemoerder\altaffilmark{1},
Moritz G\"unther\altaffilmark{1},
Paola Testa\altaffilmark{2},
Claude R. Canizares\altaffilmark{1}
}

\altaffiltext{1} {Kavli Institute for Astrophysics and Space Research, Massachusetts Institute of
Technology}
\altaffiltext{2} {Smithsonian Astrophysical Observatory, Center for Astrophysics}

\email{nss@space.mit.edu}

\begin{abstract}
The \emph{Chandra} High Energy Transmission Gratings (HETG) Orion Legacy Project (HOLP) is 
the first comprehensive set of observations of a very young massive stellar cluster which provides
high resolution X-ray spectra of very young stars over a wide 
mass range (0.7 - 2.3 \Msun). In this paper, we
focus on the six brightest X-ray sources with T Tauri stellar counterparts which
are well-characterized at optical and infra-red wavelengths. All stars show column densities
which are substantially smaller than expected from optical extinction indicating 
that the sources are located on the near side of the cluster with respect to the
observer as well as that these stars are embedded in more dusty environments.
Stellar X-ray luminosities are well above $10^{31}$ \ergsec, in some cases exceeding $10^{32}$ \ergsec for 
a substantial amount of time. The stars during these observations show no flares but are persistently
bright. The spectra can be well fit with two temperature plasma components of 10 MK and 40 MK, of which
the latter dominates the flux by a ratio 6:1 on average. The total EMs range
between 3 - 8$\times10^{54}$ cm$^{-3}$ and are comparable to active coronal sources. Limits on the
forbidden to inter-combination line ratios in the He-Like K-shell lines show that we
observe a predominantely optically thin plasma with electron densities below $10^{12}$ cm$^{-3}$. 
Observed abundances compare well with active coronal sources underlying the coronal nature
of these sources.
The surface flux in this sample of 0.6 to 2.3 \Msun classical T Tauri stars shows that coronal activity 
and possibly coronal loop size increase significantly between ages 0.1 to 10 Myrs. 
\end{abstract}

\section{Introduction\label{sec:intro}}

The Orion Nebula Cluster (ONC) is a complex stellar formation region that hosts a variety of
young stellar objects in terms of mass, age, configuration and evolutionary stages.
The cluster is part of the Orion A molecular cloud which is host to a hierarchical structure
of ongoing star formation cells~\citep{bally2000}. The part of the ONC we generally refer to 
is a somewhat older formation bubble located at the foreground of the main molecular cloud.
Its massive stars as members of the Orion Trapezium - $\theta^1$ Ori C and $\theta^2$ Ori A - are
the main sources of illumination and ionization of the Orion Nebula (M42). 
The ONC hosts one of the largest assembly of young stars in Orion with about 80$\%$ of 
its members being younger than 1 Myrs. There are over 3000 stars in the immediate
vicinity of the Orion Trapezium leading to an average stellar density of about 
250 stars per pc$^{3}$ within a radius of about 3 pc. About 1600 stars are 
optically observed and have been classified to some limited extent through spectroscopic 
and photometric measurements \citep{hillenbrand1997}. Over 2000 stars have been observed
in the IR band with 2MASS \citep{skrutskie2006} and ground based surveys ~\citep{muench2002, robberto2010}. 

X-rays from Orion and specifically the ONC were first discovered with \emph{Uhuru} \citep{giacconi1972}
as the bright X-ray source 3U0527-05. Half a decade later observations with the 
\emph{ANS} satellite suggested a more extended emission region and it was first suggested
that the X-rays are emission from coronae around T Tauri stars \citep{denboggende1978}.
This suggestion was finally confirmed by the \ein \citep{feigelson1981} and 
\ros \citep{gagne1995} X-ray Observatories. But it was the \chandra
X-ray Observatory launched in 1999 that provided the bulk of our current knowledge of X-ray detections,
identifications, and basic spectral properties. The \chandra Orion Ultradeep Project
(COUP, \citealt{feigelson2005}) detected 1600 X-ray stars and measured column densities,
source fluxes, and basic X-ray spectral and photometric parameters \citep{getman2005}.

X-ray emissions from young pre-main sequence (PMS) stars with masses 0.1 \Msun to about 2 \Msun
have 10$^5$ times the flux than stars on the main sequence. 
The bulk of emissions in PMS stars is in the optical band. The 
ratio of X-ray to bolometric luminosity in these stars lies between 
10$^{-4}$ and 10$^{-3}$, close or at the saturation threshold.
Studies of the X-ray emissions are specifically sensitive to coronal activity and
to some extent - in conjunction with UV emissions - to stellar surface shocks from 
accretion. In fact, studies within the last two decades revealed that while these X-rays
are primarily attributed to enhanced coronal activity of the star itself, 
many of these young PMS stars also generate X-rays through accretion from a 
protostellar disk \citep{kastner2002, kastner2004, schmitt2005, guenther2006,argiroffi2007}. The cases of 
TW Hya, BP Tau, V4046 Sgr, and MP Mus - all are known to be older classical T Tauri stars (CTTS) - 
are indeed quite enigmatic because here the bulk of the X-ray emission
is observed below 1.5 keV (above 8~\AA) with peculiar line ratios indicative
of accretion shock emissions. CTTS  by definition exhibit strong hydrogen Balmer (H$\alpha$) emissions
and show stong IR excess emissions up to 100 microns indicating the presence of active 
accretion disks. X-ray studies of the Taurus region also indicate that accretion
signatures are present in CTTS \citep{telleschi2007, audard2007}.  
The stars mentioned above are the best studied CTTS in X-rays today. In contrast, 
most PMS stars, CTTS and older weak-lined T Tauri stars (WTTS),
are X-ray bright due to active coronae. \citet{kastner2004} clearly
showed that coronal emissions are primarily responsible for the WTTS HID98890, \citet{telleschi2007}
showed that many CTTS have hard spectra with substantial emissions up to 10 keV,
which are definitely not due to accretion shocks. 

Evidence that most CTTS have X-ray spectra dominated by hard emission and produced by 
coronal activity comes
from very young stellar clusters and specifically the ONC as the closest massive
cluster to the Sun. Chandra observations of many young clusters in the stellar
neighborhood revealed thousands of X-ray sources, most of them CTTS, with 
X-ray spectra as hard as 2 - 3 keV \citep{feigelson2005, wolk2002, rho2004, townsley2011}.
In the ONC \citet{preibisch2005} showed that in a COUP sample of 600 X-ray sources.
which are reliably identified with optically well characterized T Tauri stars,
the plasma temperatures obtained from the X-ray spectra not only are much hotter than
usually observed in main sequence stars but also show correlations 
related to coronal activity. Most details of coronal properties in PMS stars are
still unknown as well as its evolution towards a much quieter state when these
stars reach the main sequence.  
    
In this paper we present an in-depth analysis of the X-ray properties
of six classical T Tauri stars in the ONC for which we have sufficient spectral data.
These stars are some of the brightest and most active CTTS in the COUP sample
described by \citet{preibisch2005}.

\section{Observations and Data Reduction\label{sec:obs}}

The Orion Trapezium was observed with the 
\chandra high energy transmission grating spectrocmeter HETGS~\citep{canizares2005} seventeen times
between 1999 and 2008. Table~\ref{tab:observations} summarizes the observing
parameters and exposures. While most observations were performed by 
pointing towards the brightest Orion star $\theta^1$ Ori C
(RA: 05:35:16.46, Dec: -05:23:22.85), some pointings were up to 1.5' off that direction.
Since the six PMS stars we focus on here are within 40" of $\theta^1$ Ori C, this 
created some very unfavorable combinations of off-axis angles and roll angle
overlaps. Specifically OBSIDs 4473 and
4474 are severely affected and we omit the latter from the analysis.
In OBSID 4473 we only have useful spectra for V1399 Ori.
Note, the count rate in Table~\ref{tab:observations} only shows the
cumulative rates of all six sources after spectral cleaning (see below).
While the total exposure of all observations is about 585 ks, at the end only
250 ks on average are available for each of the six stars in the sample analyzed here.  

\begin{table*}[t]
\begin{center}
\caption{Observation Log\label{tab:observations}}
\begin{tabular}{ccccc}
\hline \hline
 ObsID & Start Date & Start Time & Exposure & rate$^\dagger$\\
       &     [UT]   &   [h:m:s]  &   [ks]   & 10$^{-1}$ c/s \\
\hline
 3     & 1999 Oct 31   & 05:46:18  & 49.482 &  3.9 \\
 4     & 1999 Nov 24   & 05:36:51  & 30.914 &  1.5 \\
 2567  & 2001 Dec 28   & 12:24:53  & 46.357 &  2.9 \\
 2568  & 2002 Feb 19   & 20:28:38  & 46.334 &  2.4 \\
 4473  & 2004 Nov 3    & 01:47:00  & 49.119 &  0.3 \\
 7407  & 2006 Dec 3    & 19:06:43  & 24.633 &  1.6 \\
 7410  & 2006 Dec 6    & 12:10:32  & 13.067 &  2.4 \\
 7408  & 2006 Dec 19   & 14:16:25  & 24.862 &  2.4 \\
 7409  & 2006 Dec 23   & 00:46:36  & 27.086 &  2.1 \\
 7411  & 2007 Jul 27   & 20:40:17  & 24.631 &  0.7 \\
 7412  & 2007 Jul 28   & 06:15:04  & 24.834 &  0.9 \\
 8568  & 2007 Aug 6    & 06:53:03  & 35.860 &  1.1 \\
 8589  & 2007 Aug 8    & 21:29:30  & 50.402 &  1.5 \\
 8897  & 2007 Nov 15   & 10:02:11  & 23.644 &  1.3 \\
 8896  & 2007 Nov 30   & 21:57:29  & 22.657 &  1.3 \\
 8895  & 2007 Dec 7    & 03:13:02  & 24.851 &  1.5 \\
\hline \hline
\end{tabular}
\end{center}
$\dagger$ sum of 1st order dispersed count rates for all six sources.
\end{table*}

All observations were reprocessed using CIAO4.6 with the most recent CIAO
CALDB products. We used standard wavelength redistribution matrix files (RMF) 
and generated effective areas (ARFs) using the provided aspect solutions 
\footnote{see \url{http://asc.harvard.edu/ciao/threads/}}.
Note that for 
the HETGS spectra the RMF is fairly independent of focal plane temperature, however
the order sorting has to be adjusted to the different pulse height distributions
of the CCD at -110 C in OBSIDs 3 and 4. Here we followed the steps applied in the data reduction
for the Trapezium stars described in previous analyses using these data sets
~\citep{schulz2000, schulz2006}. For all the HETGS observations we 
generated spectra and analysis products for the medium energy gratings (MEG) +1 and -1
orders, as well as for the high energy gratings (HEG) +1 and -1 orders. 

\smallskip
\includegraphics[angle=0,width=8.5cm]{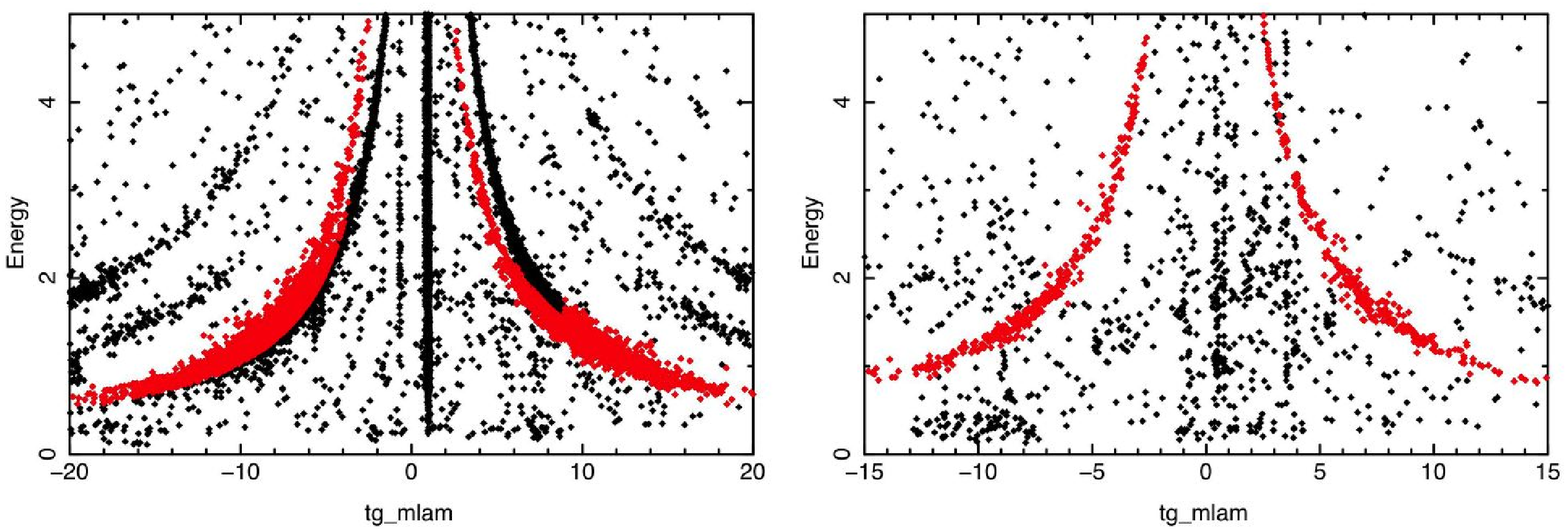}
\figcaption{Plots of grating dispersion (tg$\_$mlam) versus CCD array pulse height
channels (energy) for one OBSID for MEG (left) and HEG (right). The source dispersion
of the bright star MT Ori (red) overlaps significantly with another source
(black) rendering the MEG portion unuseable.\label{fig:banana} }
\smallskip

A crowded field of stars and observations at
various different roll and off-axis angles create additional challenges in the 
proper spectral data extraction. In order to clean the data there are several steps
to consider. These are described in detail in \citet{huenemoerder2009} for the 
extraction of the young and bright intermediate Orion Trapezium binary $\theta^1$ Ori E.
The first one concerns the various off-axis angles of the sources with respect
to the telescopes optical axis. The HETG spectral resolution degrades significantly
once a source is well beyond 1' from the optical axis of the telescope. 

As a next step the orders from the grating dispersion are sorted
by the pulse heights (PHA) of the CCDs. However sources with rolls causing overlaps
of their dispersed spectra from other sources spatially will suffer from sharing PHAs 
of dispersed orders from these sources. The result cannot be resolved and these
data have to be discarded. Depending on roll angle, single grating arms are affected
differently and each arm for each source has to be investigated separately. Note that
these overlaps can be caused by any of the sources in the field including the bright
sources of the Orion Trapezium ~\citep{schulz2001, schulz2003}. Figure~\ref{fig:banana} 
shows an example of such a case in MT Ori data where PHA orders of two different sources overlap.
In all of these cases the grating orders cannot be used.
On average these occurrences remove about 40$\%$ of the available exposure for 
each source.

\smallskip
\begin{figure*}
\includegraphics[angle=0,width=16.5cm]{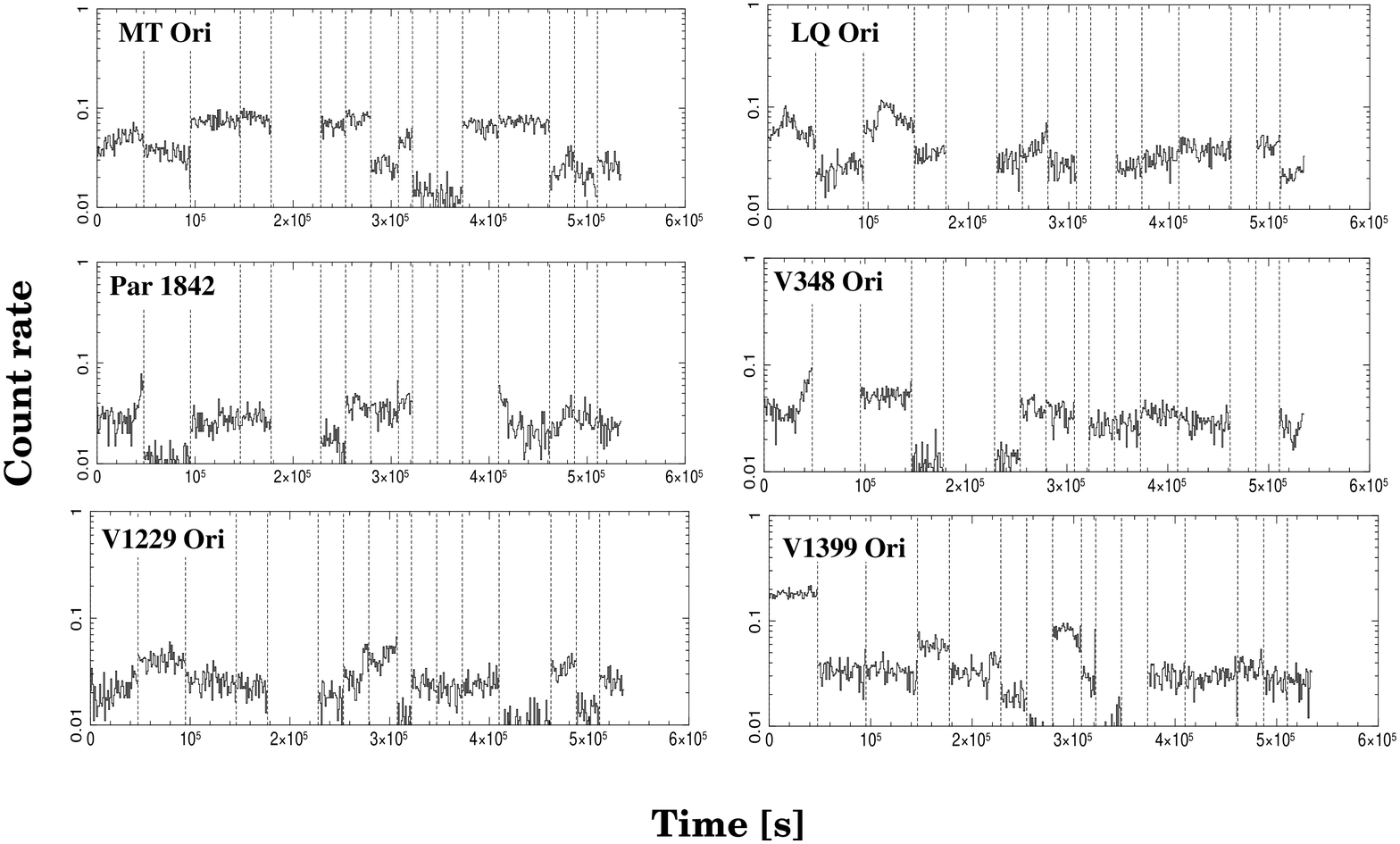}
\figcaption{These lightcurves were compiled from HETG 1st order fluxes where available. We
abstained from using zero order data because of pile up issues. Observations are separated
by the dotted lines, The UT times are listed in Tab.~\ref{tab:observations}. \label{fig:lightcurves} }
\end{figure*}
\smallskip

In a separate step coincidences of other source zero orders located on HETG dispersion arms have
to be eliminated as they appear as spurious line emissions in the dispersion.
Many of these coincidences can be allocated using the source locations from
\citet{getman2005}. However, since there are considerable numbers of X-ray
flaring sources in the ONC, a final visual inspection of each dispersion arm
was warranted. In this respect any line feature in the four grating arms
for each source had to be consistently present, otherwise we suppressed that
particular bandpass in the dispersion. The rates in Table~\ref{tab:observations} are 
calculated from the dispersed spectra after all cleaning steps were completed.
 
\section{General Source Properties\label{sec:general}}

\subsection{Optical Properties\label{sec:optprop}}

Physical parameters of very young PMS stars are hard to determine as at that young stage
much of the central star is still enshrouded in natal circumdisk material. This means
that accurate determinations of extinction and accretion luminosities are difficult. 
A summary of possible biases in the age determination of young PMS stars can be
found in ~\citet{soderblom2014}. The
standard for ONC members was set by a study by ~\citet{hillenbrand1997}, which was
exclusively used in the COUP analysis ~\citep{preibisch2005}. A more recent
improved multi-color optical survey provided updated extinction values and effective
temperatures for a variety of ONC stars~\citep{dario2010} (see also ~\citealt{herczeg2014}). 
The study by ~\citet{dario2010} produced a new
Hertzsprung-Russell diagram of the ONC population. Using the tracks calculated by
~\citet{siess2000}, this study provided new values for physical parameters such as
A$_V$, accretion luminosity fractions L$_{acc}$/L$_{tot}$, masses M$_{\star}$ and 
stellar age a$_{star}$. The values for the six stars in our sample are summarized
in Table~\ref{tab:params}. For the stellar ages we also include the
~\citet{hillenbrand1997} results, a$_{h97}$, for reference purposes.

\begin{table*}
\begin{center}
\begin{small}
\caption{Low-Mass PMS Stars: Optical Properties\label{tab:params}}
\begin{tabular}{lcccccccc}
\hline \hline
 Name  &  log T$_{eff}$ &  log L$_{\star}$ &  AV & log L$_{acc}$/L$_{\star}$ &  M$_{\star}$ & R$_{\star}$ & log a$_{\star}$ & log a$_{h97}$ \\
       &     [K]        &    \Lsun         &     &                           &  \Msun      &  \Rsun  &    [yr]         &     [yr]      \\
 \hline 
 MT Ori   &  3.66 & 1.08 &  1.89  &  -0.76 &  1.99 & 8.2 & 5.6 & 6.2 \\
 LQ Ori   &  3.60 & 0.82 &  0.46  &  -2.42 &  0.70 & 7.7 & 4.8 & 6.0 \\
 Par1842  &  3.75 & 0.64 &  2.84  &  -0.37 &  1.56 & 3.3 & 7.0 & 6.4 \\
 V1399Ori &  3.71 & 0.81 &  1.61  &  -0.74 &  2.28 & 4.6 & 6.4 & 6.4 \\
 V1229 Ori &  3.72 & 0.86 &  2.98  &  -0.45 &  2.22 & 5.0 & 6.5 & 6.5 \\
 V348 Ori &  3.72 & 0.91 &  2.97  &  -0.68 &  2.33 & 4.7 & 6.5 & 6.3 \\
\hline \hline
\end{tabular}
\end{small}
\end{center}
\end{table*}

The effective temperatures for all six stars range from $\sim$ 4000 K (LQ Ori) to 
about 5300 K (V348 Ori), which puts these stars currently into the G/K type category.
Except for LQ Ori, the stars still produce a significant amount of luminosity from
active accretion with fractions between $\sim$ 20 and 40 $\%$. Masses and ages
have ben determined from PMS tracks provided by the model from ~\citet{siess2000}, which
gives a mass range between 0.7 \Msun and 2.3 \Msun and ages between 7$\times10^4$ yr and 
1$\times10^7$ yr. Most accurate knowledge of these two quantities is essential because 
they provide the stellar radii we need to interprete the coronal emissions results.

\subsection{X-Ray Properties from the COUP Study\label{sec:coup}}

The sources in this sample are the X-ray brightest stars within a 1.5' radius
around the Orion Trapezium. The source selection is purely based on their presence
in the optimal field of view (FOV) of the Chandra HETG spectrometer.
The X-ray properties of these stars as determined from the COUP dataset 
are listed in Table~\ref{tab:COUP} and more of 
their global properties are shown in ~\citet{getman2005}  
and ~\citet{preibisch2005}. The COUP properties of our sample 
are affected by photon pileup up to $100\%$ and in 
all the cases mitigating steps had to be taken in the COUP analysis.
Absorption columns were found to be significantly lower than in the
bulk of X-ray sources ranging between 1$\times10^{20}$ cm$^{-2}$ and 
2$\times10^{21}$ cm$^{-2}$. The spectra were fitted by a two temperture
plasma model yielding average temperatures of 8$\times10^6$ K and 
1.9$\times10^7$ K, up to over a factor two lower than most other COUP
souces~\citep{preibisch2005}. Emissivities are 2.6$\times10^{54}$ cm$^{-3}$
and 5.5$\times10^{54}$ cm$^{-3}$ and comparable to the entire COUP sample.

\begin{table*}[b]
\begin{center}
\begin{small}
\caption{Low-Mass PMS Stars: COUP Properties\label{tab:COUP}}
\begin{tabular}{*{11}{c}}
\hline \hline
 Name & COUP$\#$ & Spec. Type & RA & Dec & log N$_{h}$ & kT$_{1}$ & kT$_{2}$ & log EM$_1$ & log EM$_2$ &log L$_x$ \\
      &          &            &    &     & (cm$^{-2}$) &  (keV)   &  (keV)   & (cm$^{-3}$)& (cm$^{-3}$) & (\ergsec)\\ 
\hline
MT Ori    & 932  & K2-K4& 05 35 17.94 & -05 22 45.5 & 21.17 & 0.76 & 1.63 & 54.8 & 55.1 & 32.18\\
LQ Ori    & 394  & K2V  & 05 35 10.73 & -05 23 44.6 & 20.00 & 0.68 & 1.71 & 54.0 & 54.1 & 31.32\\
Par 1842  & 689  & G7-G8& 05 35 15.23 & -05 22 55.7 & 20.85 & 0.64 & 1.64 & 54.2 & 54.8 & 31.82\\
V1399 Ori & 1130 & G8-K0& 05 35 21.04 & -05 23 49.0 & 21.28 & 0.71 & 1.51 & 54.4 & 54.4 & 31.67\\
V1229 Ori  & 965  & G8-K0& 05 35 18.35 & -05 22 37.4 & 21.25 & 0.64 & 1.63 & 54.4 & 54.8 & 31.89\\
V348 Ori  & 724  & G8-K0& 05 35 15.62 & -05 22 56.4 & 20.46 & 0.80 & 1.45 & 54.3 & 54.6 & 31.73\\
\hline \hline
\end{tabular}
\end{small}
\end{center}
\end{table*}

\subsection{X-Ray Fluxes from HETG\label{sec:hetg}}

We fit the spectra with a two component plasma model using the
Astrophysical Plasma Emission Database APED~\footnote{http://www.atomdb.org~\citep{smith2001}}
\citep{foster2012}. In
this first fit we do not constrain parameters such as abundance or absorption,
but model the continuum as accurately as possible. This first fit already provides
reliable spectral fluxes and luminosities. The values shown in Table~\ref{tab:HETG} are
the stellar surface fluxes and surface integrated luminosities for a distance
of 450 pc. We use the latter distance in order to directly compare
the results to published COUP data. The actual distance toward the ONC has recently
been determined to 412 pc~\citep{reid2009}. As done in \citet{preibisch2005} the stellar X-ray flux
is defined as a surface flux and is the luminosity divided by the stellar surface
area. 
   
\begin{table*}
\begin{center}
\caption{HETG Spectral Parameters of 2 Temperature APED fits\label{tab:HETG}}
\vskip 4pt
\begin{tabular}{lccccccc}
\hline
\hline
 Name &  N$_{H}$ & T$_1$ & T$_2$ & EM$_1$ & EM$_2$    & F$_{x}$ & L$_{x}$ \\
 &(1) &[MK]&[MK]& (2)& (2)&(3) & (4) \\
\hline
 & & & & & & &\\
MT Ori &  1.35 $^{  0.14 }_{  0.14 }$ & 11.00 $^{ 0.39 }_{ 0.47 }$ & 39.24 $^{ 1.95 }_{ 1.95 }$ &  1.2 $^{  0.1 }_{  0.1 }$ &  6.6 $^{  0.2 }_{  0.2 }$ & 20.2 $^{  1.2 }_{  1.2 }$ &  8.4  \\
LQ Ori &  0.00 $^{  0.15 }_{  0.00 }$ & 9.39 $^{ 0.27 }_{ 0.27 }$ & 39.10 $^{ 2.38 }_{ 2.38 }$ &  1.4 $^{  0.1 }_{  0.1 }$ &  3.4 $^{  0.1 }_{  0.1 }$ & 13.8 $^{  0.8 }_{  0.8 }$ &  5.0  \\
Par 1842 &  0.51 $^{  0.28 }_{  0.17 }$ & 11.09 $^{ 0.54 }_{ 0.60 }$ & 37.50 $^{ 2.92 }_{ 2.92 }$ &  0.7 $^{  0.1 }_{  0.1 }$ &  2.3 $^{  0.1 }_{  0.1 }$ & 46.0 $^{  2.8 }_{  2.8 }$ &  3.1  \\
V348 Ori &  0.83 $^{  0.18 }_{  0.18 }$ & 11.04 $^{ 0.46 }_{ 0.53 }$ & 45.23 $^{ 2.95 }_{ 2.95 }$ &  0.9 $^{  0.0 }_{  0.1 }$ &  3.3 $^{  0.1 }_{  0.1 }$ & 35.4 $^{  2.1 }_{  2.1 }$ &  4.6  \\
V1229 Ori &  0.84 $^{  0.20 }_{  0.19 }$ & 10.34 $^{ 0.52 }_{ 0.88 }$ & 36.41 $^{ 1.94 }_{ 1.94 }$ &  0.8 $^{  0.1 }_{  0.1 }$ &  3.0 $^{  0.2 }_{  0.1 }$ & 25.5 $^{  1.5 }_{  1.5 }$ &  3.9  \\
V1399 Ori &  0.72 $^{  0.18 }_{  0.18 }$ & 11.21 $^{ 0.40 }_{ 0.37 }$ & 37.96 $^{ 2.79 }_{ 2.79 }$ &  1.0 $^{  0.1 }_{  0.1 }$ &  3.0 $^{  0.1 }_{  0.1 }$ & 31.8 $^{  1.9 }_{  1.9 }$ &  4.3  \\
& & & & & &  \\
\hline
\end{tabular}
\end{center}
(1) 10$^{21}$ cm$^{-2}$
(2) 10$^{54}$ cm$^{-3}$
(3) 10$^6$ erg cm$^{-2}$ s$^{-1}$
(4) 10$^31$ erg s$^{-1}$
\end{table*}

The most luminous PMS star in the sample is MT Ori with 8.4$\times10^{31}$ \ergsec, the least luminous
star is 3.1$\times10^{31}$ \ergsec, but all luminosities are lower than  
deduced in the COUP analysis.
The ratio of X-ray to bolometric luminosity is log L$_x$/L$_{bol}$ = -2.77$\pm$0.08 and very
similar for all sources. Detected X-ray fluxes range between 1$\times10^{-12}$ \ergcm and 3$\times10^{-12}$ \ergcm.
For the luminosities we corrected these fluxes for the absorption column. The stellar X-ray surface fluxes
(F$_x$ in Tab.~\ref{tab:HETG}) were then determined by dividing these luminosities by the stellar surface area using stellar
radii associated with the model parameters in Table~\ref{tab:params} and \citet{siess2000} tracks.  

\subsection{Variability\label{sec:variab}}

In order to determine source variability we use only the rate deduced from 
the first order HETG spectra in order to avoid any biases from pileup in the 
zero order point sources. The light curves are shown in Fig.~\ref{fig:lightcurves}.
The 1st order count rates are plotted for all observation segments, which
in the light curves are separated by dotted lines. The observing times of
each segment thus appear on a continuous scale even though the segments are
weeks, months, or even years apart. Some segments do not show data, this does not
mean that there is no source flux but merely that we do not have valid data
coverage after the confusion analysis.
The total observing time is 5.85$\times10^5$ s.

We do not observe any significant flaring activity in each of the observing segments. 
The sources appear
persistent in most observing segments with limited short term flux variations. 
Notable rate variations of a few factors on time scales of a few hours appear
in LQ Ori and V1399 Ori. The larger changes in flux appear in between observations.
The largest change between observations 
appears in V1399 Ori with almost a factor 10 difference in mean flux per observation. 
There might be flare onsets in the the first segments of Par 1842 and V348 Ori,
but these appear at the end of the observations and are thus hard to classify
as a flare. Note that even in the COUP sample these sources do not have 
a flare history	~\citep{getman2008}.

\begin{figure*}
\includegraphics[angle=0,width=17.5cm]{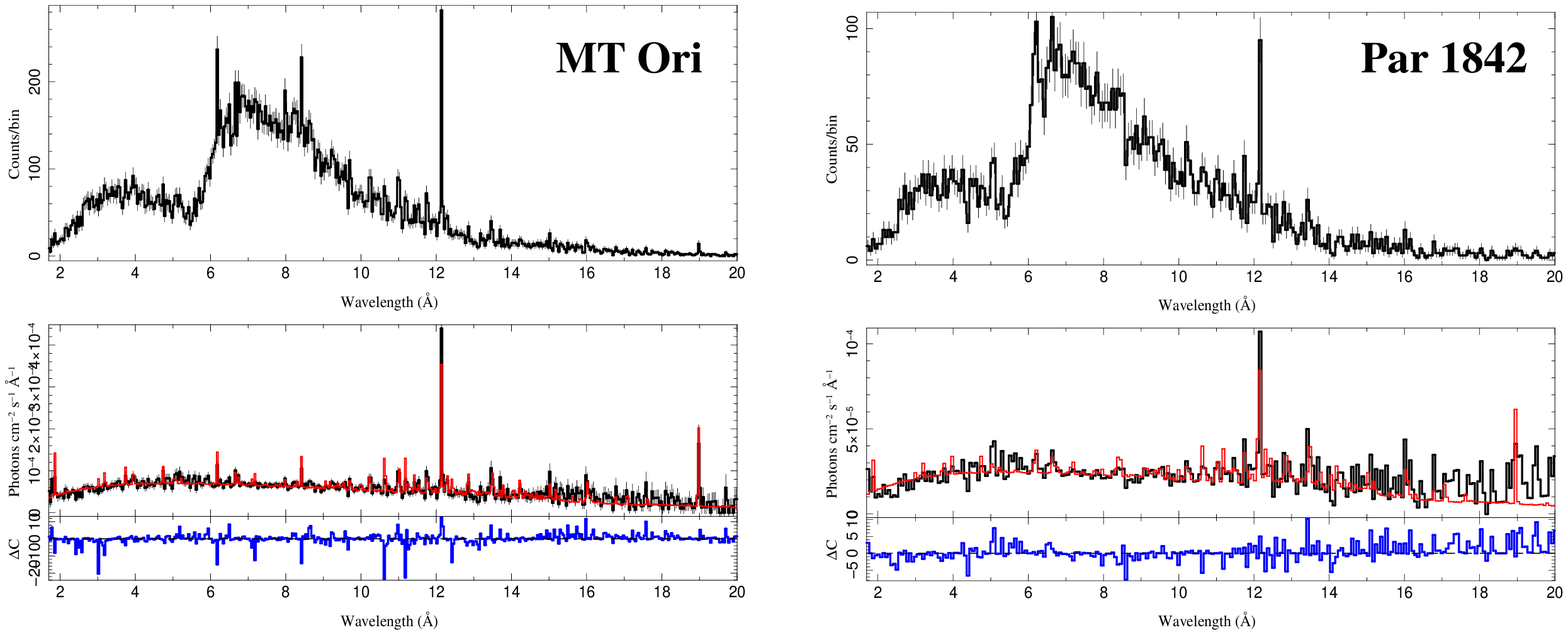}
\figcaption{Two temperature APED fits of the combined 1st order HETG spectra of
for the brightest source MT Ori and the weakest source Par 1842. \label{fig:spectra1}}
\end{figure*}

\section{Spectral Analysis\label{sec:analysis}}

The modeling of the spectra and the X-ray line emission is done in several steps.
The models used in the analysis are two temperature APED model spectra for the
overall spectral analysis and local continua and Gaussian functions for the 
emission line analysis. A two temperature model became necessary and sufficient
since a one temperature unconstrained fit produced strong residuals and 
some unfeasible abundance parameters and models with more temperature components did not
appear statistically significant. Hence the two temperature APED models
are the most consistent way to fit the spectra of all six sources. Spectra and 
fit models are shown in Figs.~\ref{fig:spectra1} and ~\ref{fig:spectra2}.

\smallskip
\begin{center}
\includegraphics[angle=0,width=8.0cm]{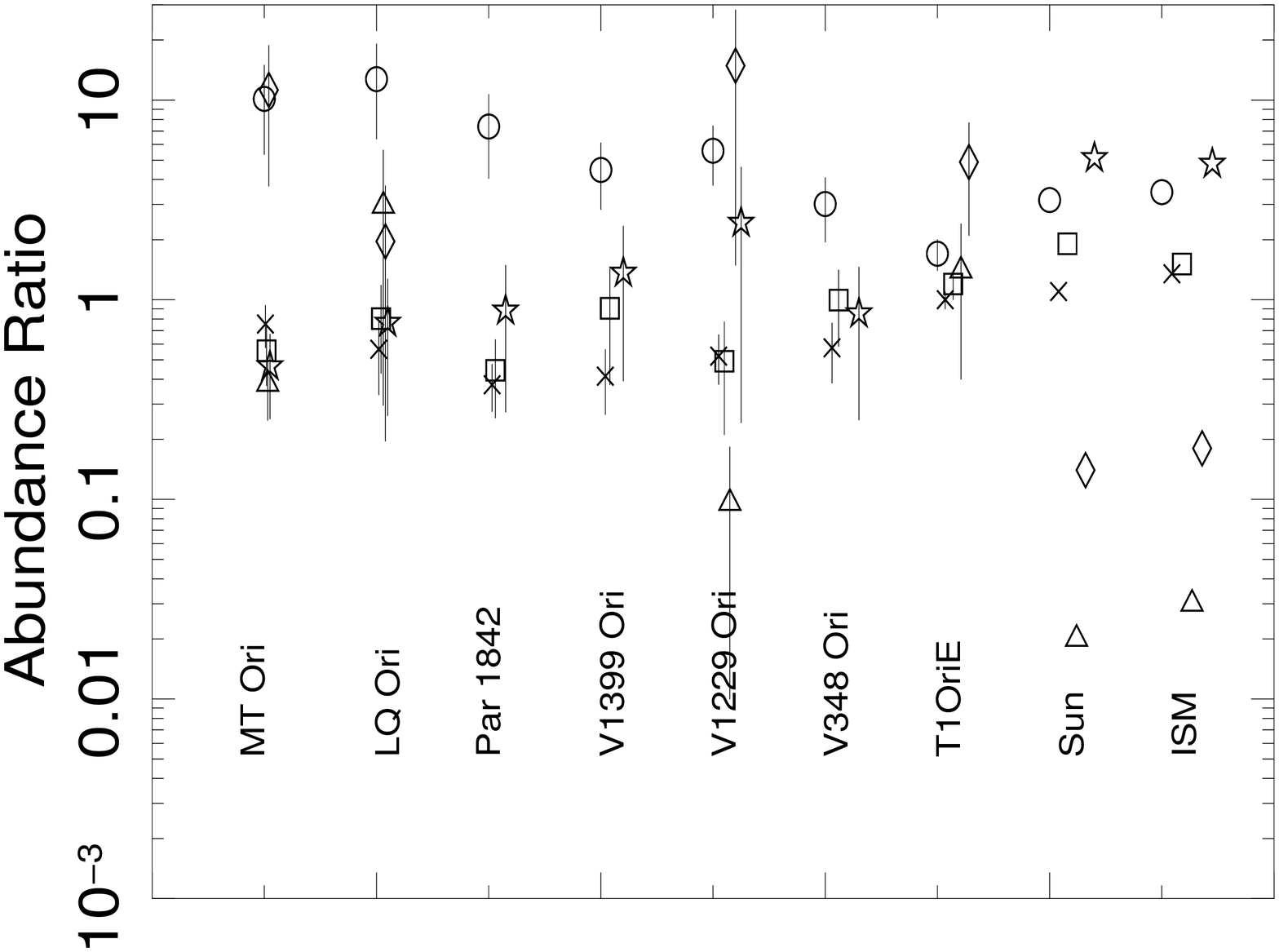}
\figcaption{Unfolded two temperature APED fits of the combined 1st order HETG spectra of
four of the Orion sources. \label{fig:spectra2}}
\end{center}
\smallskip

\subsection{Abundance Line Ratios\label{sec:abundratios}}

Most of the power of the HETG spectra comes from the detection of distinct X-ray
line emission. The measurement of K-shell line properties puts significant
constraints on the physical nature of the emitting plasma. Collisional plasmas
at temperatures above 10$^6$ K emit K-shell lines from all major cosmic abundant
elements in the available wavelength range from 1.7 to 25 \AA\ of the HETG spectrometer.
In each spectrum we detect over 100 X-ray lines of which about 70$\%$ are very
weak Fe lines. In all the spectra the lines appear unresolved and 
)match the expected locations (see also ~\citet{huenemoerder2003} with no apparent shifts
and line broadenings.

\begin{table*}
\input{tab1.tex}
\end{table*}

The HETG band covers most abundant
non-iron H-like lines and He-like line triplets except for C and in most cases N. 
Measured line fluxes are shown in Table~\ref{tab:linefits}. Many lines, specifically 
for the He-like triplets appear spurious or we only see upper limits. The upper
limit line flux in our sample is about 2$\times 10^{-7}$ photons cm$^{-2}$ s$^{-1}$.
For high Z He-like triplets such as Ar, S, and Si we only see a blend and significantly
model the resonance and forbidden line components. The detection of higher Z lines from 
Ca and Fe are more rare.

Ratios of the line fluxes
are used to determine a common abundance distribution with a first goal to constrain the APED 
plasma model fits. Here we utilize temperature insensitive abundance ratios
as used by ~\citet{drake2005, drake2005a, liefke2008}, and further developed by \citet{huenemoerder2013}, which read as

\begin{equation} 
{A_i\over A_j} = {a_o} {{F_{i,H} + a_1 F_{i,He r}}\over{F_{j,H} + a_2 F_{j,He r}}},
\label{eq:ratio}
\end{equation}

where $A_i, A_j$ are abundances of element i and j, $F_{i,H}, F_{j,H}$ the line
fluxes (in photons cm$^{-2}$ s$^{-1}$) of the corresponding H-like ions, 
$F_{i,He r}, F_{j,He r}$ the resonance
line components of the corresponding He-like line triplets. 
The coefficients a$_o$, a$_1$, and a$_2$ were determined
from APED and are listed in ~\citet{huenemoerder2013} assuming abundances of ~\citet{anders1989}.


The temperature insensitive abundance ratios we calculate from this procedure
are listed in Table~\ref{tab:ratios} and plotted in Fig.~\ref{fig:orionratios}.
There are more recent and updated distributions such as the one from ~\citet{grevesse1998} and
~\citet{asplund2009}. The differences, however, are well within the uncertainties of the X-ray abundance
ratios and we did not re-scale the values to the most recent distribution~\citep{asplund2009}.

\smallskip
\includegraphics[angle=0,width=8.5cm, angle=0]{fig5.eps}
\figcaption{Measured abundance ratios for Orion sources. The symbols mean:
circle(Ne/Mg); cross (Mg/Si); square (Si/S); trangle (S/O); diamond (Ne/O); star (S/Ar).
Added are the values from the sun~\citep{grevesse1998} and the ISM~\citep{wilms2000}
\label{fig:orionratios}}
\smallskip

The average value of these ratios for each ratio type was the basis to construct a set
of abundances and associated uncertainties. These were used to constrain the 
abundance parameters for O, Ne, Mg, Si, S, and Ar during the
APED plasma fits below. We also compare these ratios to the values expected from the Sun
~\citep{grevesse1998} and the ISM~\citep{wilms2000}. We observe that some values such as
Mg/Si or Ne/Mg or Si/S align fairly well with the Sun and the ISM, S/O or Ne/O or S/Ar
are significantly different indicating anomlies with the oxygen and/or neon abundances. 
In our case \ovii was only poorly detected, however.
The abundance of Fe was not included in this procedure for the lack of K-shell lines
and ratio coefficients. Here we only can adjust the abundance during the APED fits.

\begin{table*}
\begin{center}
\begin{small}
\caption{Temperature insensitive line ratio with uncertainties\label{tab:ratios}}
\begin{tabular}{lcccccc}
\hline \hline
 Name & Ne/Mg & Mg/Si & Si/S & S/O & Ne/O & S/Ar  \\
      &       &       &       &    &      & \\
\hline
  MT Ori   & 10.2$\pm$4.8 & 0.7$\pm$0.2 & 0.6$\pm$0.2 & 0.4$\pm$1.3&11.3$\pm$7.5 & 0.5$\pm$0.2\\
  LQ Ori   & 12.7$\pm$6.4 & 0.6$\pm$0.2 & 0.8$\pm$0.4 & 3.0$\pm$2.7& 2.0$\pm$1.8 & 0.8$\pm$0.5\\
  Par 1842 &  7.4$\pm$3.3 & 0.4$\pm$0.1 & 0.4$\pm$0.2 &   --          &    --    & 0.9$\pm$0.6\\
  V348 Ori &  4.5$\pm$1.6 & 0.4$\pm$0.3 & 0.9$\pm$0.5 &   --          &    --    & 1.4$\pm$1.0\\
  V1229 Ori &  5.6$\pm$1.8 & 0.5$\pm$0.2 & 0.5$\pm$0.3 & 0.1$\pm$0.1&14.9$\pm$13.4& 2.4$\pm$2.2  \\
  V1399 Ori&  3.0$\pm$1.1 & 0.6$\pm$0.2 & 1.0$\pm$0.4 &   --          &    --    & 0.9$\pm$0.6\\
$\theta^1$ Ori E& 1.7$\pm$0.3& 1.0$\pm$0.1    & 1.2$\pm$0.2     & 1.4$\pm$1.0   & 4.9$\pm$2.8   &      \\
  Sun      &  3.16           &  1.10          & 1.91            & 0.02          & 0.14          & 4.12 \\
  ISM      &  3.46           &  1.35          & 1.51            & 0.03          & 0.18          & 4.79 \\
\hline \hline
\end{tabular}
\end{small}
\end{center}
\end{table*}

\subsection{R-Ratios from He-like Triplets\label{sec:helike}}

In collisionally ionized plasmas the He-like line triplets are sensitive
to density and external UV fields. He-like line triplets consist of resonance (r),
intercombination (i), and forbidden (f) line components. At increasing collisional plasma
densities the meta-stable f-line component depopulates into the i-line components.
A similar effect appears by the influence of far-UV photons matching the energy difference
between the f- and i-lines.

In the absence of a strong UV radiation field,
the flux ratio ($R=f/i$) of the two lines is sensitive to the density of the plasma. Coronal plasmas are considered
low density plasmas with electron/ion densities below the order of 10$^{11}$ cm$^{-3}$,
while plasmas in accretion columns can reach 10$^{14}$ cm$^{-3}$~\citep{kastner2002}. He-like triplets of
O, Ne, and Mg ions are sensitive to this range. In general, R-ratios significantly below 4.5 indicate    
densities significantly exceeding 10$^{10}$ cm$^{-3}$~\citep{testa2004}. For neon, ratios significantly below 3.8 
indicate densities significantly exceeding 10$^{11}$ cm$^{-3}$. And last, but not least, for 
magnesium, ratios significantly below 2.8 indicate densities significantly exceeding 10$^{12}$ cm$^{-3}$.  
However, even though many R-ratios in our sources are below these limits, in nearly all but a few cases
the result is not very significant because of the low statistic in the lines. At this stage the
uncertainties do not allow us to discern from the low density limit.

\smallskip
\includegraphics[angle=0,width=8.5cm]{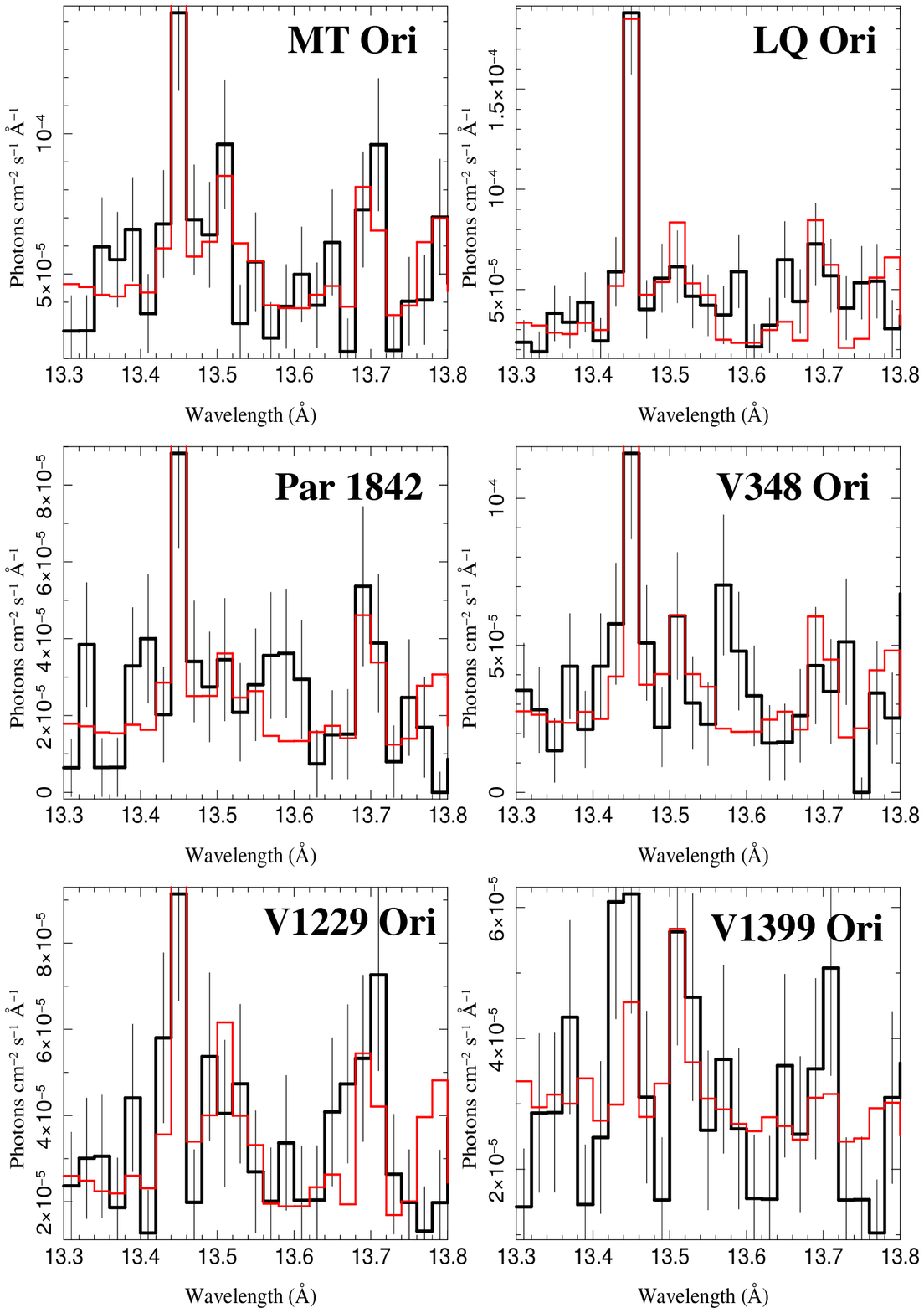}
\figcaption{APED plasma fits of the \neix tiplet regions in the six Orion sources~\label{fig:neix}.}
\smallskip

Figure~\ref{fig:neix} shows the \neix\ triplet region for all sources. Of all the lower Z triplets
the \neix\ triplets appear more prominently in our spectra than the \mgxi\ and \ovii\ triplets,
also likely due to a neon overabundance.
Weak He-like line strengths are likely due to the high plasma temperatures these spectra,
which favor resonance line strengths. 
This can be seen in Figure~\ref{fig:neix}, where in all sources the \neix\ resonance line
dominates the bandpass. The \neix\ bandpass in Figure~\ref{fig:neix} was fit with the lower temperature APED
plasma component and the results shows that a plasma of low density ($< 10^{11}$ cm$^{-3}$) fits the 
region well within the limited statistics. 

\subsection{Column Densities\label{sec:columns}}

The broadband plasma fits produce unusually low column densities, some with values 
well below expectations from the COUP analysis, suggested values from
observed optical extinction, and suggestions from previous analysis
of the massive stars in the ONC ~\citep{schulz2001, gagne2005} which
suggest columns of at least 2.9$\times10^{21}$ cm$^{-2}$. Columns can be estimated
from optical extinction via~\citep{predehl1995}

\smallskip
\includegraphics[angle=0,width=6.5cm]{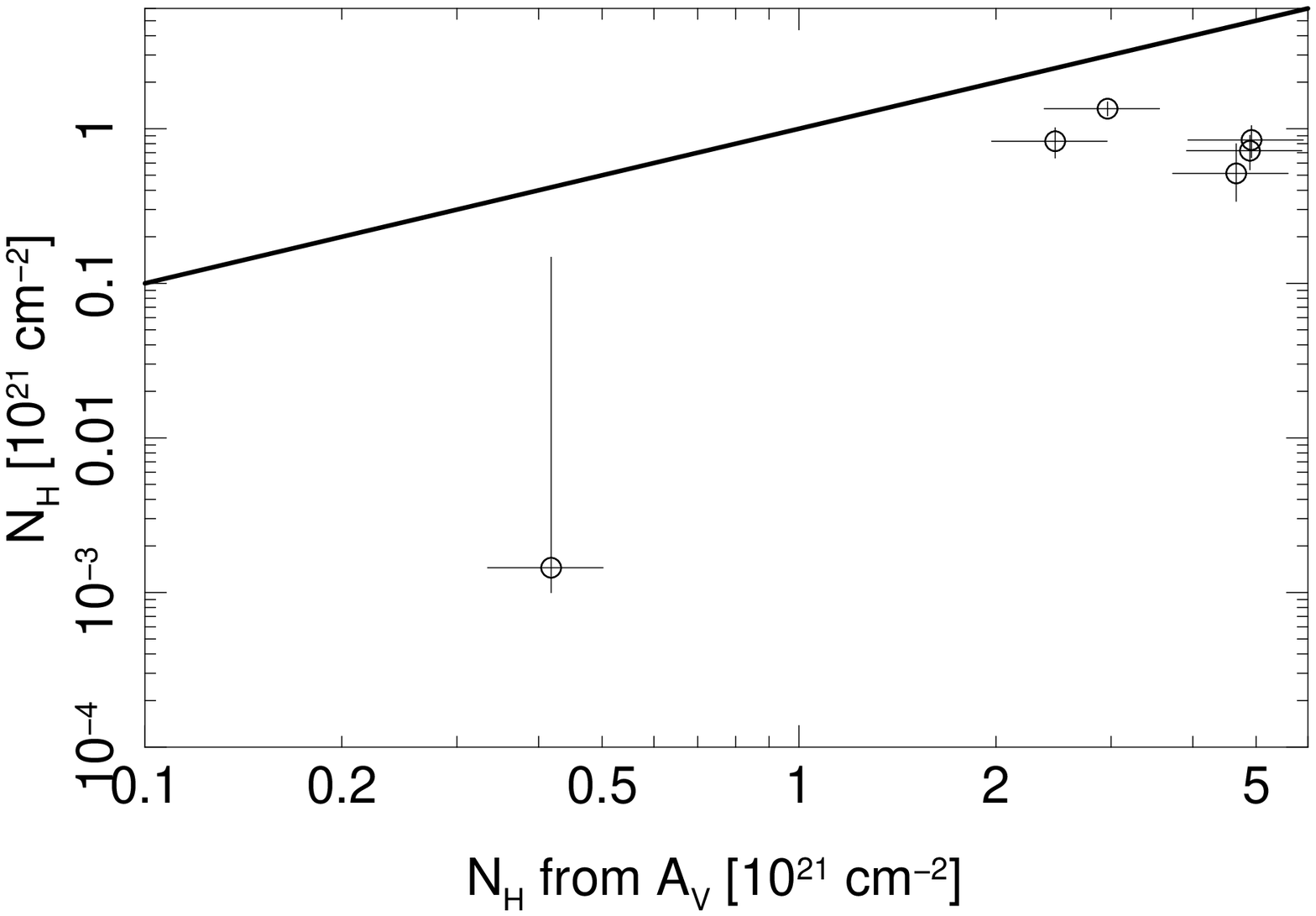}
\figcaption{The column densities as predicted from optical extinction A$_V$ versus
the final columns found in the plasma model fits. The bold line indicates the expectation
of the two values being the same.\label{fig:columns}}
\smallskip

\begin{equation}
A_V = 0,56\times N_H - 0.23,
\label{avcol}
\end{equation}

where A$_V$ is the extinction (see Table~\ref{tab:params}) and N$_H$ is the X-ray column in
units of 10$^{21}$ cm$^{-2}$. Figure~\ref{fig:columns} shows the columns from optical extinction A$_V$
versus the measured X-ray columns. Here we used the columns from the APED model fits.
These columns are significantly lower than the one expected from extinction.    

We have also determined line-of-sight absorption from line fluxes of
approximately known intrinsic ratios.  
Once emission lines are from a single ion, the ratios are determined by
the atomic data, and to a weaker extent, on the plasma temperature
distribution. If these intrinsic ratios are known, then the observed
ratios are determined by the absorption function, which then can be
inverted to give the column density.  We used the H-like Lyman-like series with 
\oviii\ $\alpha$ ($\lambda 18.973\AA$) and $\beta$
($\lambda 16.006\AA$) lines.  The Ly-$\beta$ like
line, however, is blended with \fexviii\ $\lambda 16.004\AA$
(lower and upper transition configurations of $2p^43s$ -- $2p^5$).

Using another \fexviii\ line at $14.208\AA$ (a doublet; $2p^43d$
-- $2p^5$), we can break the degeneracy~\citep{testa2004a, testa2007}. 
We have three measured
fluxes comprised of 4 lines, with two known ratios (an \fexviii\
ratio, and the \oviii\ ratio).  The other unknown is the absorbing
column.  This constitutes five equations with five unknowns, so we can
solve for $N_\mathrm{H}$.  The abundances are not a parameter since we
are ultimately using the ratios of identical ions.

The system is non-linear (since the absorption function is
exponential), so we solved this numerically.  We used the
``{\tt phabs}'' function from {\tt XSPEC}~\citep{arnaud1996} and the emissivities
of AtomDB~\citep{foster2012}.
We show an example fit to the relevant MT~Ori spectral region in
Figure~\ref{fig:mtoriabsfit}.

\smallskip
\includegraphics[width=8.5cm]{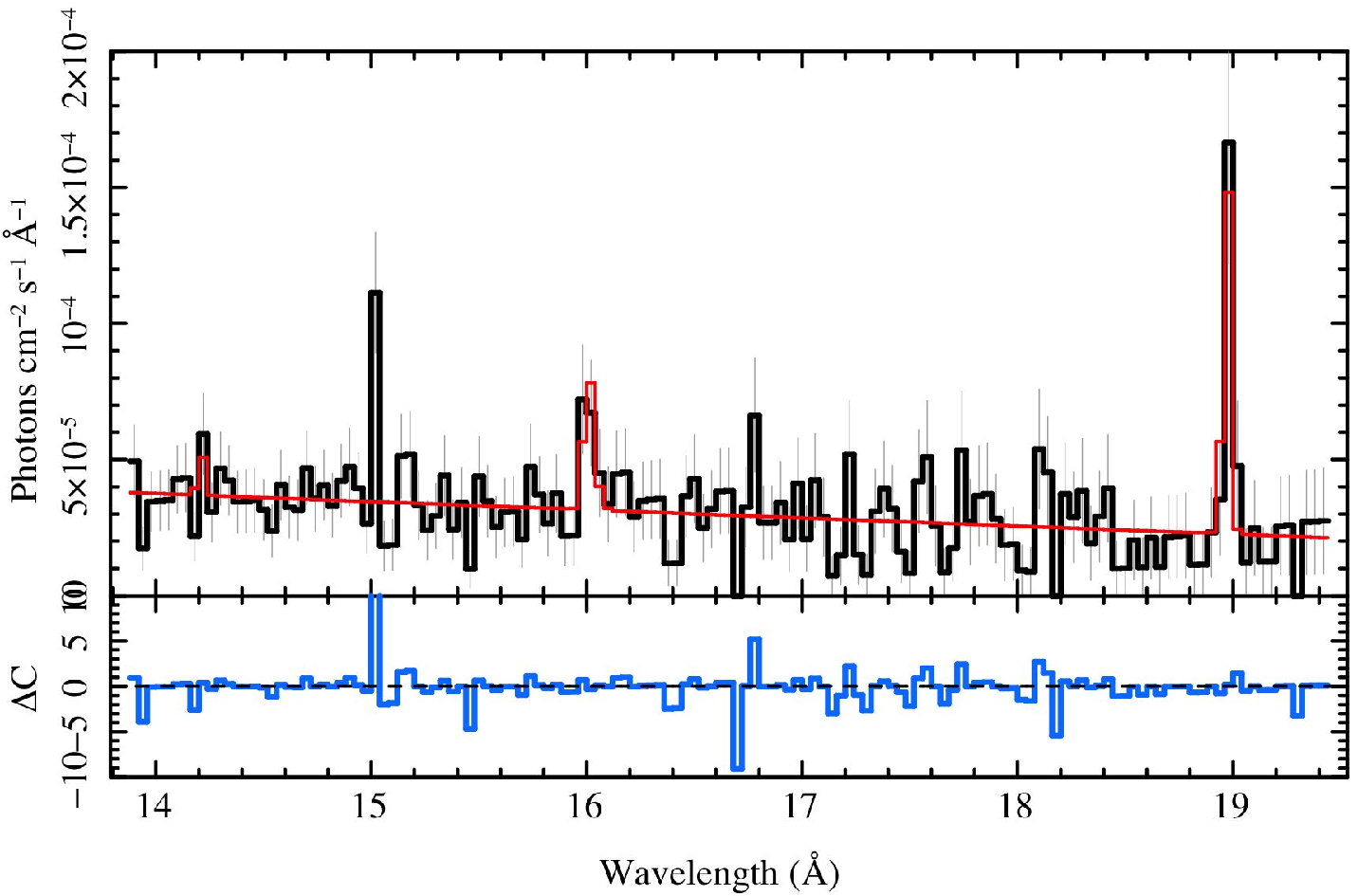}
  \figcaption{The MT~Ori region from which we derived the absorbing
    column from emission line ratios.  The fit function was a constant
  plus four Gaussian functions.  The lines of interest are \fexviii\
  $\lambda 14.208\AA$, \oviii\ $\lambda 18.967\AA$, and the
  unresolved blend of \fexviii\ $\lambda 16.004\AA$ and
  \oviii $\lambda 16.006\AA$.  We also included a resolved blend
  of \fexviii\ $\lambda 16.071\AA$ in order to get a better flux
  of the unresolved blend.\label{fig:mtoriabsfit}}
\smallskip

Given the weakness of the $14.208$ \AA\ line, we performed a
Monte-Carlo analysis in which we perturbed the line fluxes according
to their uncertainties (assuming Gaussian distributions). 
Distributions of the column density were then obtained for $10^4$ realizations. We did
this for the PMS stars in this sample as well as for the more massive ONC
stars, and for two bright, unabsorbed objects, $\sigma\,$Gem and
HR~1099~\citep{huenemoerder2013}. We also added a distributon for the sum of the Orion
PMS sample stars. The Orion PMS stars consistently yield a
most probable value of
$N_\mathrm{H}\sim0.2\times10^{22}\,\mathrm{cm^{-2}}$.  The ``control''
stars ($\sigma\,$Gem and HR~1099) yielded much lower values of about
$0.03$--$0.05$ which is consistent as they are expected to be
consistent with almost $0.0$.
Figure~\ref{fig:nhfits} shows the distributions obtained for the set
of Orion stars analyzed. The peaks of line ratio distribution is mostly consistent
with the columns from the broadband analysis, LQ Ori appears a little bit higher but
the 90$\%$ uncertainties align well with the propabilites just below the peak.
We conclude that the columns in the Orion stars in this sample are indeed very low.

\smallskip
  \includegraphics[width=8.5cm]{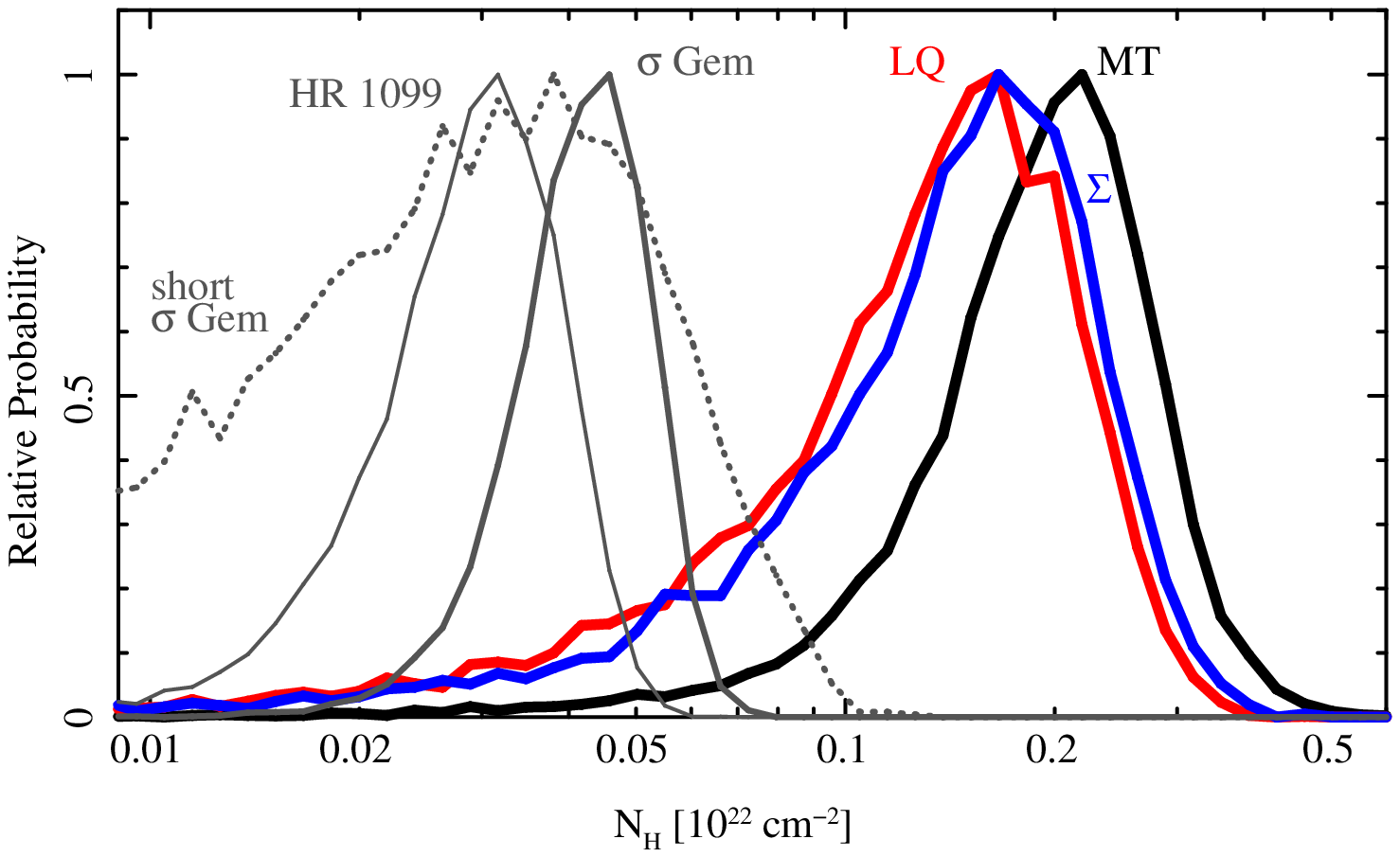}
  \figcaption{The fits to the collection of stars, as determined from
    $10^4$ Monte-Carlo realizations for each star.\label{fig:nhfits}}
\smallskip

\subsection{APED Fits: Temperatures and Abundance\label{sec:aped}}

Broadband fits of all spectra were performed in several steps. The first step involved
a restricted fit using a first set of abundance settings obtained from the temperature insensitive
line ratios. The uncertainties in the set are large
and allowed the fit considerable freedom. In consecutive steps we manually adjusted single elements during the fits
in order to minimize residuals which appeared in the He-like and H-like lines. For the 
final distibution we calculated $90\%$ uncertainty limits for all values.
The final abundance
distribution for the fits then yields the following values with respect to solar~\citep{anders1989}:
O (0.195$\pm0.38$), Ne (0.718$\pm0.23$), Mg (0.085$\pm0.034$), Si (0.153$\pm0.07$), S (0.225$\pm0.120$), 
Ar (0.524$\pm0.220$), Fe (0.058$\pm0.012$). 

Multi temperature components did not produce
statistically superior solutions with respect to two temperature models. 
The final two temperature fits are shown in Fig.~\ref{fig:spectra1} and 
Fig.~\ref{fig:spectra2} and the parameters are
shown in Tab.~\ref{tab:HETG}. The scatter on most spectra is very large above
20~\AA\ and for the final APED fits we limited the bandpass to 1.8 and 20~\AA.
The data bins were grouped to the size of 0.04~\AA\ in the brightest and 0.08~\AA\ in the 
weakest source for display purposes. For the fits, each spectrum had 941 independent data bins,
each bin had the size of one MEG resolution element (0.021~\AA). 

The resulting X-ray plasma properties we measure in our sample of 
six sources, i.e. emissivities and temperatures, 
generally agree well with the overall COUP sample trends in ~\citet{preibisch2005}. 
This is, however, not true for the source parameters measured in COUP~\citep{getman2005}.
A comparison of Table~\ref{tab:COUP}
with Table~\ref{tab:HETG} shows that fluxes and emissivities are overestimated and
temperatures significantly lower in the COUP  data, which likely adds to the scatter in the
COUP results. 

\section{Results and Discussion\label{sec:disc}}

\subsection{Flux and Luminosity\label{sec:flum}}

Even though the limited number of sources in this sample does not allow
for any statistical trend analysis, we can point out
some differences and similarities between these sources. 
There are some differences in the
overall luminosity, the weakest source Par 1842 has 3.1$\times10^{31}$ \ergsec, the 
brightest source MT Ori has 8.4$\times10^{31}$ \ergsec. The spread increases when
we compare the surface flux, i.e. the X-ray luminosity divided by the stellar
surface area. Figure~\ref{fig:sflux} shows the surface flux with respect to the 
listed ages from~\citet{dario2010}, which indicates that in the age range 
up to 10 MK we observe a substantial increase in coronal activity. This trend
is implicitely reflected when estimating coronal loop sizes, in this case
using $em \sim \rho^2 r_l^3$, where $em$ is the volume emission measure,
$\rho$ a coronal density of 10$^{11}$ cm$^{-3}$, and $r_l$ the loop size.
This is a very simplistic estimate, we here also just assume a filling factor 
of unity and similar densities, but it shows that these sizes also increase with age, 
The fact that surface normalized X-ray fluxes increase with age for 
CTTS in the ONC has been reported by COUP, but for lower mass stars~\citep{preibisch2005a}
up to 1 \Msun but with no significant effect for 1 - 2 \Msun stars. Here stars range
from 0.7 to 2.3 \Msun and the effect is highly significant. One should take the
case of LQ Ori with caution, however, its age determination of $< 0.1$ Myr is unrealistically
low and not supported by models, but based on the data~\citet{dario2010} it is 
still likely younger than MT Ori.  

\subsection{Column Densities and Extinction\label{sec:extinct}}

Except for LQ Ori all sources in the sample are affected by substantial
amounts of extinction (Tab.~\ref{tab:params}), which is not unusual and even expected towards
the ONC. In \ros studies~\citet{predehl1995} investigated the
relation of X-ray column density and optical extinction and found a
quite robust relationship (Eq.~\ref{avcol}) specifically valid for
columns above 10$^{21}$ cm$^{-2}$. However it was developed for
long range line of sights with gas and dust distributed over vast
distances. In the case of Orion we have the situation that extinction
is local to the dust in the ONC and its sources, while the X-ray column
reflects absorption towards Orion and is likely gas dominated. In other words,
there must be a different offset than in Eq.~\ref{avcol} due to local dust, which is
not observed in the X-ray column. 
Such an offset in A$_V$ is clearly visible in Fig.~\ref{fig:columns}. 

However, such interpretations have to be viewed with caution.
In star forming regions the extinction is indeed local and the extinction
law might differ from the properties in the interstellar medium (ISM). In
particular, the properties of dust grains, which cause the optical extinction,
are expected to change as the star forming regions evolves. In the initial cool
starless cores temperatures are much lower and densities much higher than in
the ISM, favoring dust formation, while later the intense radiation from the
newly-formed stars will evaporate smaller dust grains.
Observationally, one way to probe this regime is to compare the X-ray
absorption $N_H$ with the optical extinction $A_V$. X-ray absorption is caused
by the total column density (gas and dust) of heavy ions and is expressed as
the equivalent hydrogen column density, assuming some standard set of abundances.
On the other hand, the optical extinction is dominated by the dust column density
and also depends on the dust grain size distribution.
\citet{vuong2003} compared $N_H$ and $A_V$ for six star forming regions, including the ONC.
For the $\rho$~Oph star forming region they find an $N_H/A_V$ significantly below the ISM 
value and they interpret this as a sign that the cloud material in $\rho$~Oph has a lower 
metal abundance than the ISM, consistent with recent solar abundance measurements.
Alternatively, grain growth can increase the amount of extinction per unit mass
until the grains reach about 1 $\mu$m in size \citep{ormel2011}.
\citet{vuong2003} have only a few data points with a large scatter in the ONC, 
which are roughly consistent with an ISM-like $N_H/A_V$ ratio, but in our data (Fig. 5) we
see that all targets have an $N_H/A_V$ ratio significantly below the ISM value,
in line with the \citet{vuong2003} result for $\rho$~Oph.
In fact, low $N_H/A_V$ values are seen in several low-mass star forming regions,
e.g.\ NGC 1333 \citep{winston2010}, and IRAC 20050+2720 \citep{guenther2012}. Confusingly,
previous studies of high-mass star forming regions (RCW 38: \citet{wolk2006}; 
RCW 108: \citet{wolk2008})
tend to find ISM-like $N_H/A_V$ ratios. It should be noted, however, that the
studies cited above all use CCD spectra to measure $N_H$ and that there
often is an ambiguity between the amount of cool plasma and the absorbing column
density, such that a high column density and a large amount of cool plasma or a
low column density with no absorption provide and equally good fit. In our
study we break this degeneracy through the use of high-resolution grating
spectroscopy. \citet{wolk2008} used a similar approach for older,
but still accreting CTTS in low-mass star forming regions and found consistently
high $N_H/A_V$ ratios, indicating gas-rich material. They speculated
that some of their sources could be seen through the accretion column, which would be
dust-depleted due to the stellar irradiation.
 
We are left with a complex picture where the dust content, grain growth, and
abundance in the circumstellar matter and the ISM all influence the observed
$N_H/A_V$ ratio and their respective influence in the sightline to a
specific
target cannot be disentangled. For our six targets in the ONC, the total
extinction is low, so there cannot be much cloud material in the line of
sight. Given that the objects are all very young, some circumstellar matter
with evolved dust grains and possibly a low metal abundance can explain the
observed $N_H/A_V$ ratios.

\smallskip
\includegraphics[angle=0,width=8.0cm]{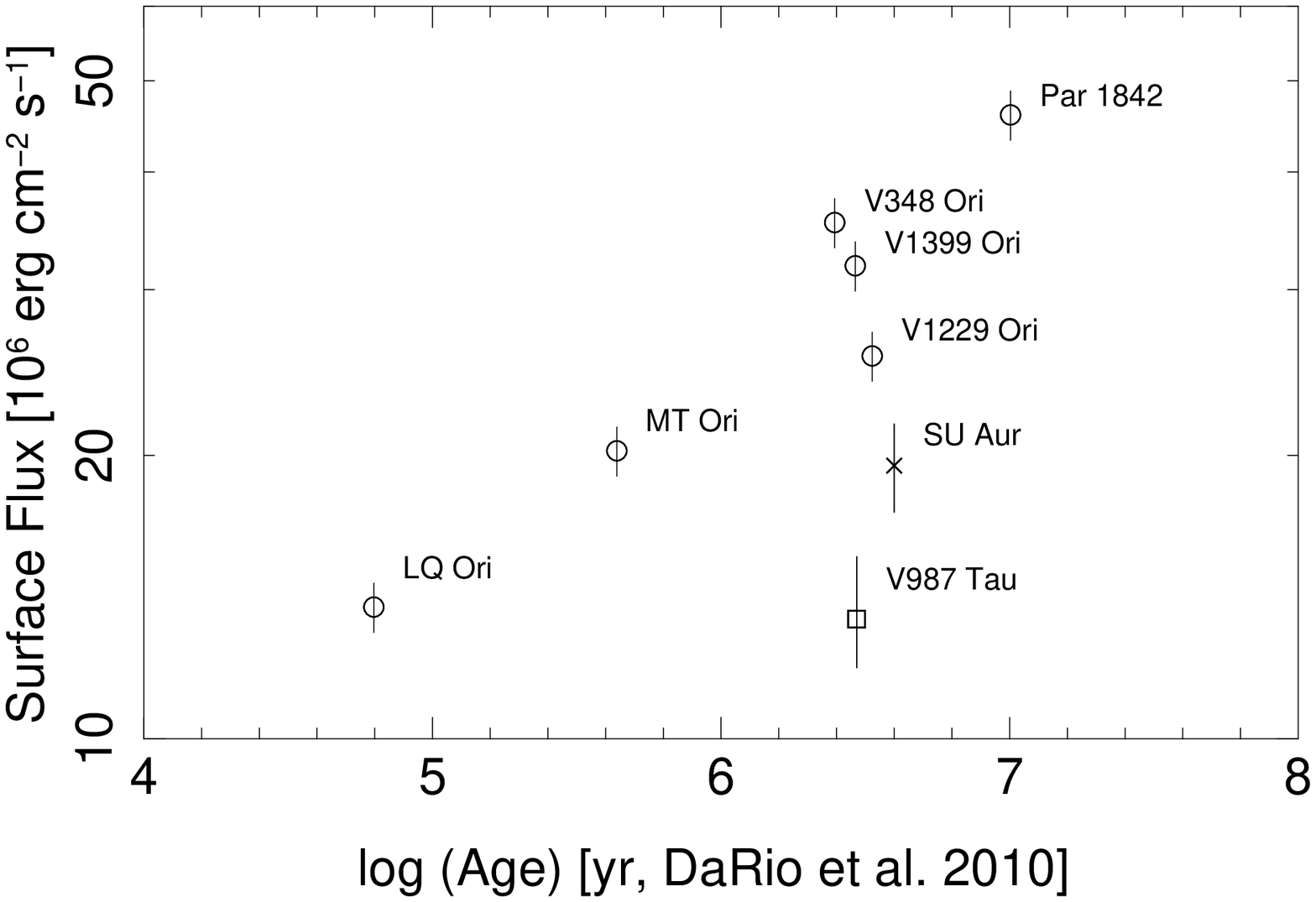}
\figcaption{The stellar surface X-ray flux versus stellar age.\label{fig:sflux}}
\smallskip

\subsection{Plasma Densities and Accretion\label{sec:plasma}}

The fact that we cannot use observed R-ratios because of our limiting
statistics at this stage is unfortunate, but one also has to put
their value for our star sample into perspective.
At these high plasma temperatures even for the lower temperature component,
the He-like triplets are not only difficult to observe, but we would not
see much contribution from accretion shock plasmas in these sources. These
stars have simply not contracted enough to support sufficiently high infall
velocities, which range from as low as 150 km s$^{-1}$ in LQ Ori to
about 350 km s$^{-1}$ in Par 1842. For our predictions we use
the values in Table~\ref{tab:params} and a simple model from \citet{calvet1998}.
For example, in order to ionize the plasma up to \mgxi\ temperatures above about 2 MK
are necessary, which requires infall velocities above 800 km s$^{-1}$. This
cannot be the case here and we would not expect signatures from accretion
activity.

Even though our fits of the \neix\ regions are consistent with a low density
plasma, we are statistically limited to rule out higher density plasmas.
These might be observed in specific coronal environments at the basis of strong coronal loops
involving small emitting volumes and high temperatures~\citep{sanz2003, testa2004} or
coronal heating of dense and flared inner accretion disk regions~\citep{drake2010,
drake2009, dullemond2010, dullemond2007}. Magnetic confinement as outlined
by~\citet{gagne2005} might be a possibility in these type of interactions as well.

In the following subsections we dicuss how these issues fare with
individual sources and how these properties compare with other
PMS stars and results from the COUP sample.

\subsection{Individual Sources\label{sec:sources}}

{\bf MT Ori} is the X-ray brightest low-mass PMS star in our sample and
with a surface temperature of 4600 K currently a K2 - K4 spectral type
\citep{dario2010}. The latter study determines a contribution of about
13$\%$ of the total luminosity from accretion. 
At an age of about 0.5 Myr (see Tab.~\ref{tab:params}), one might expect
more from accretion in a 2 \Msun CTTS. The ratio of the emission measures of the
two temperature components yields a fraction of 18$\%$ for the lower temperature
component. An accretion fraction would be consistent with  
the low optical accretion luminosity, however, the estimated infall velocity of 250 km s$^{-1}$
is too low to allow for K-shell ionization of matter with atomic number larger
than \ovii. We should thus not expect low R-ratios from an infall accretion shock
for \neix\ and \mgxi. 

The total X-ray luminosity is very
high leading to a log L$_{x}$/L$_{bol}$ of about -2.7, above the 
saturation limit of -3 for fast rotators (e.g. ~\citet{vilhu1984, vilhu1987})
and well above the median value
of -3.7 for TTS with unknown periods in COUP data. However, the surface
X-ray flux is the 2nd lowest in the sample, due to the fact that the stellar
surface area is the largest in the sample. The lightcurve shows the
source persistently bright at levels varying by a factor of 8. 
From the average total emissivity we project average coronal loop
sizes of about 10$\%$ of the stellar radius assuming coronal densities
of 10$^{11}$ cm$^{-3}$. ~\citet{getman2008} found a typical flare
in the COUP dataset and estimate loop sizes of about 1.1$\times10^{11}$ cm,
which is only 30$\%$ larger. Note, the estimated
loop sizes use a model by ~\citet{reale1998}, which depends on the square root
of the observed flare temperature. However, because of pileup problems (see Sect.~\ref{sec:aped})
this temperature is about a factor two lower in the COUP dataset and the true sizes are larger by a factor
$\sqrt{2.3}$.

{\bf LQ Ori} is the second X-ray brightest low-mass PMS star in our sample
and classified as a K2 V dwarf. \citet{dario2010} find less than 1$\%$ 
of the optical luminosity due to accretion, which for a 0.6 \Msun star
at an age of the order of 0.1 Myrs is highly unusual. Ages this low are
not supported by most model calculations and we expect LQ Ori to be
somewhat older. The fraction of the low temperature component is almost 40$\%$ 
indicating the potential for some accretion to produce X-rays.
However, the case for X-rays from accretion in LQ Ori is even worse than for MT Ori as predicted
infall velocities do not even favor much ionization up to \ovii. 
LQ Ori also does not appear to be a particularly dusty environment, 
extinction is extremely low
in this star and this is backed up by a very low column density allowing to
record even a strong N~VII L$_{\alpha}$ line. log L$_{x}$/L$_{bol}$ is slightly lower than
the one in MT Ori, but still well above -3. But similar to MT Ori we estimate
an average coronal loop size of about 10$\%$ of the stellar radius.
A incomplete flare in the COUP data provided about 2.5 times the size~\citep{getman2008}.

{\bf Par 1842} is the oldest star in the sample according to \citet{dario2010}
with an age of 10 Myr. Its determined luminosity fraction from accretion is the highest in
the sample.
The low temperature component produces about 30$\%$ of the total emission measure,
and while models predict infall velocities of the order of 350 km s$^{-1}$ allowing for
both, ionizations up to \ovii\ and \neix\, but not to \mgxi\ are feasible.
The X-ray luminosity is the lowest of all stars, but its surface flux the highest,
indicating that this star is actually coronally most active. Consequently from the 
total emission measure we estimate average loop sizes of over 20$\%$ of the stellar radius.
One flare in the COUP data, which was classified as a slow rise event, 
gave a loop limit over 10 times larger (5$\times10^{11}$ cm) indicating that this 
flare was a giant event~\citep{getman2008},

{\bf V348 Ori, V1229 Ori, V1399 Ori } have very similar optical properties in terms
of surface temperature, mass, radius, and age~\citep{dario2010}. Overall this
is also true for their X-ray properties. The optical luminosity from accretion
(see Table~\ref{tab:params}), however,
varies by about a factor two between V1229 Ori, V348 Ori and V1399 Ori 
likely indicating that accretion rates vary by factors independent of system
parameters. R-ratios in all three sources show reduced values 
and the presence of cooler dense plasma from accretion
cannot be ruled out. The case here is similar to the one for Par 1842, 
ionizations in the accretion shock plasma might favor \ovii\ or \neix\ but not \mgxi. 
log L$_{x}$/L$_{bol}$ 
is also above -3 and average coronal loop sizes are of the order of 20$\%$ of the stellar radius.
No flares were detected in the COUP sample~\citep{getman2008}.

\subsection{Comparison with other T Tauri stars\label{sec:ttauri}}

There are few detailed studies of young CTTS exhibiting dominant
coronal properties other than the Orion studies. High resolution X-ray studies dealing with 
young PMS stars at various evolutionary stages have been conducted 
by \citet{audard2005} for V987 Tau, RY Tau, and LkCa 21, or by \citet{robrade2006}
for the four CTTS BP Tau, CR Cha, SU Aur, and TW Hya. 
Perhaps the most prominent non-Orion example from these samples is
SU Aur. It was found that coronal emission
is the dominant source of X-ray activity in this star with most emissions
between 20 - 50 MK and the coolest component at around 8 MK. Due to absorption,
the O~VII triplet is not observed. ~\citet{shukla2009} find R-ratios of Mg and Ne
that limit plasma densities to less than 10$^{12}$ cm$^{-3}$ and conclude a dominant
coronal nature. This is very similar to what we find for the ONC sources, except
for the fact that ONC sources appear significantly more luminous.  
Consequently SU Aur appears below the Orion source data points in Fig.~\ref{fig:sflux}.

Most PMS coronal sources are classified as weak lined T Tauri Stars (WTTS). Here
the IR-excess from a prominent accretion disk has vanished and we do not
expect any accretion activity. X-ray emissions are considered
to entirely be of some form of coronal origin. Prime examples are 
HD 98800 in the TW Hya association~\citep{kastner2004} and V987 Tau in the 
Taurus-Auriga region~\citep{strassmeier1998}.
These non-flaring states of WTTS are well described by either a one temperature component
or a weak hot and dominant very hot component. The case of V987 Tau is of interest
because its age is 2 -3 Myr ~\citep{furlan2006, damiani1995}.  
~\citet{andrews2005} found a disk mass of $< 0.0004$ \Msun 
indicating that accretion has ceased in this system. \citet{shukla2009} 
estimate R-ratios of Mg and Ne that limit plasma densities to less than 10$^{12}$ cm$^{-3}$, 
however for a very weak cool component. In Fig.~\ref{fig:sflux} V987 Tau appears
even further below the Orion stars and SU Aur~\citep{siwak2011, audard2005}, which 
is likely a consequence of its much earlier spectral type.


\subsection{Abundances in the ONC\label{sec:abund}}

The ONC as an ensemble of coeval stars allows to study details in the chemical composition
of these stars and the cluster. Coronal chemical compositions are often different from stellar
photospheric and interstellar compositions~\citep{feldman1992, testa2010}. 
Most of these differences relate to depletion effects
with respect to the first ionization potential (FIP). The solar corona has more elements
with low FIPs such as Mg, Si, Fe, while coronally active stars seem to have 
more higher FIP elements such as O, Ne, Ar. Figure~\ref{fig:studies}
compares the average abundance ratios of the Orion stars with various studies. Solar photospheric
ratios~\citep{grevesse1998, asplund2009} seem to be quite in agreement with abundances
from B-stars ~\citep{asplund2009} and B-stars in Orion~\citep{nieva2011}. A study of
the upper solar atmosphere at 10$^6$ K seems to also differ significantly from the Orion ratios.
The ratios for the Orion stars seem to agree with 
most of these studies only for the Ne/Mg, Mg/Si, and Si/S ratios, while significant differences
can be found with respect to the S/O, Ne/O, and S/Ar ratios. This indicates that 
there are likely discrepancies with at least the neon and oxygen abundance. Enhanced 
neon has been observed in X-ray spectra of other stars than the Sun (see \citealt{drake2005a}
and reference therein). Besides enhanced neon, the very high Ne/O ratio could also indicate
depletion of oxygen. Given that column densities are not extraordinarily 
high and that we observe the \nvii\ line in some cases shows that there could be a depletion
of oxgyen in the Orion spectra. This can have several reasons ranging from the fact that
spectra at these high temperatures mostly generate H-like \oviii\ lines, but also due to the large
extinction values, which indictate that these CTTS are likely quite dusty and oxygen is depleted
into dust grains~\citep{drake2005}.

\smallskip
\vbox{
\includegraphics[angle=0,width=8.0cm]{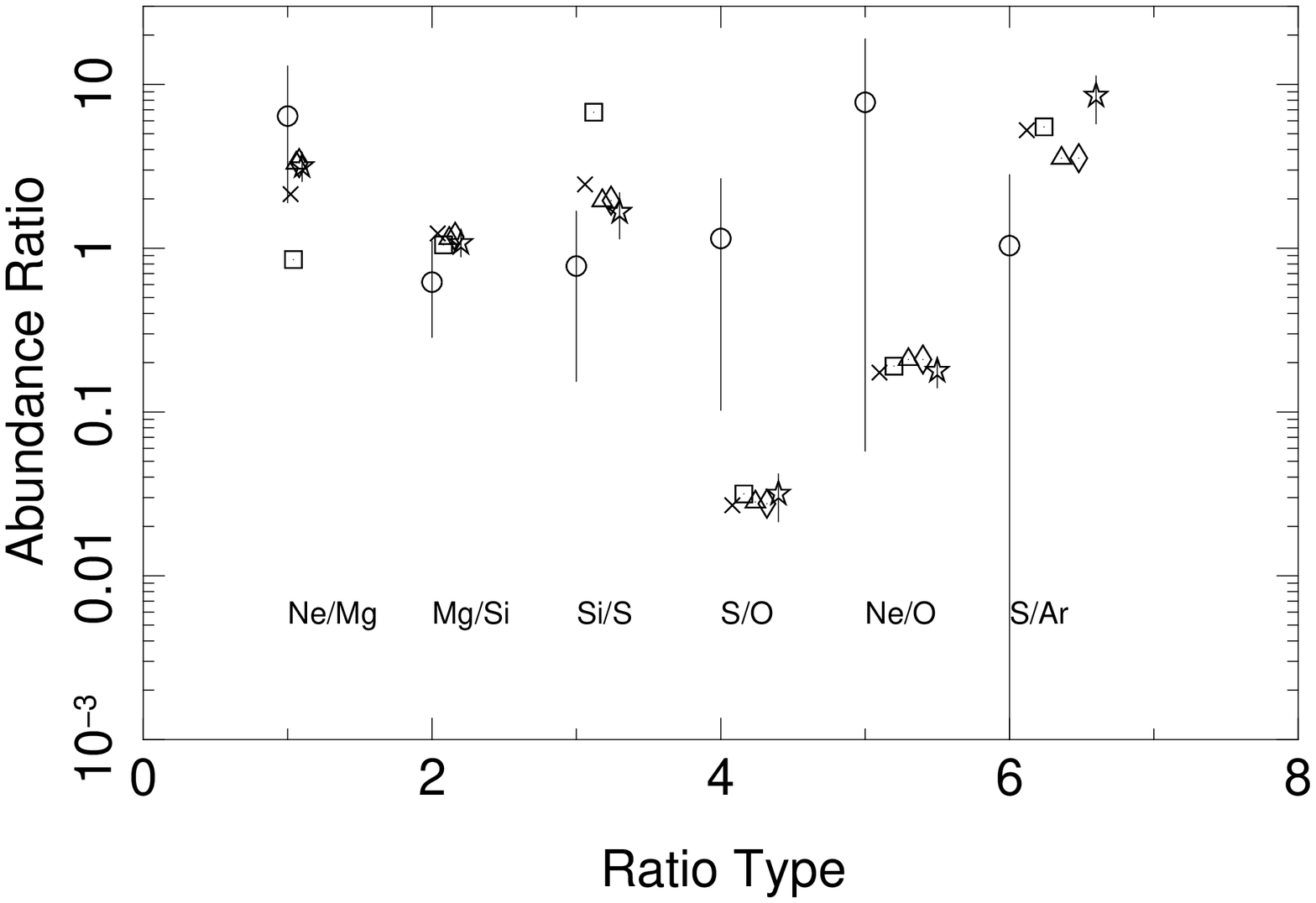}
\figcaption{The average abundance ratios for our Orion sources (circles) in comparison
with various studies. The symbols mean:
star (\citet{grevesse1998}); cross (\citet{asplund2009}); triangle (B-stars from \citet{asplund2009}); 
diamond (\citet{nieva2011}, S and Ar from \citet{asplund2009}); square (\citet{feldman2003}).
\label{fig:studies}}
}
\smallskip

Figure~\ref{fig:mdwarfs} shows a similar comparison but now involving stellar active coronae.
Here we used two studies, one for a set of fast rotating M dwarfs~\citep{liefke2008} and 
two active coronal K stars ~\citep{huenemoerder2013}. Even though for some ratios the uncertainties
are still large and there are large variations within Orion sources there is a very good
agreement with ratios of coronally active sources. Even the Ne/O ratio with its quite large
uncertainties seems more consistent with coronal souces~\citep{huenemoerder2013, drake2005}
than with solar values~\citep{asplund2009}. The comparison of 
Fig.~\ref{fig:studies} and Fig~\ref{fig:mdwarfs} thus highlights that elemental abundances in stellar
photospheres differ from stellar coronae. This difference in abundances is still an unsolved problem. 

The measured abundance ratios also show that they are insensitive to a range of
very high coronal temperatures but sensitive to a more global temperature range. However,
exactly this difference between photospheric and coronal abundances can work in favor as a 
diagnostic probe for coronal structures (see \citealt{laming2012, laming2009, sanz2003}, testa2015).
In our Orion sample we are still far from doing so given the current statistical limitations
in the X-ray line fluxes. 

Abundances of iron could not be included in the line ratio study but from the spectral plasma
fits they appear consistently low in all Orion sources. Low Fe abundances are also found
in active stellar coronae~\citep{huenemoerder2013} but also in X-ray spectra of the 
Orion Trapesium such as the intermediate mass binary $\theta^1$ Ori E~\citep{huenemoerder2013} and the massive 
stars $\theta^1$ Ori C and $\theta^2$ Ori A~\citep{schulz2003,schulz2006}. However, in
all these cases that observed plasma is likely of also coronal origin. Other studies
of Orion~\citep{nieva2011} and the solar neighborhood do not support such an iron 
abundance deficit, pointing once again to chemical fractionation effects in the outer stellar atmosphere.

\smallskip
\includegraphics[angle=0,width=8.0cm]{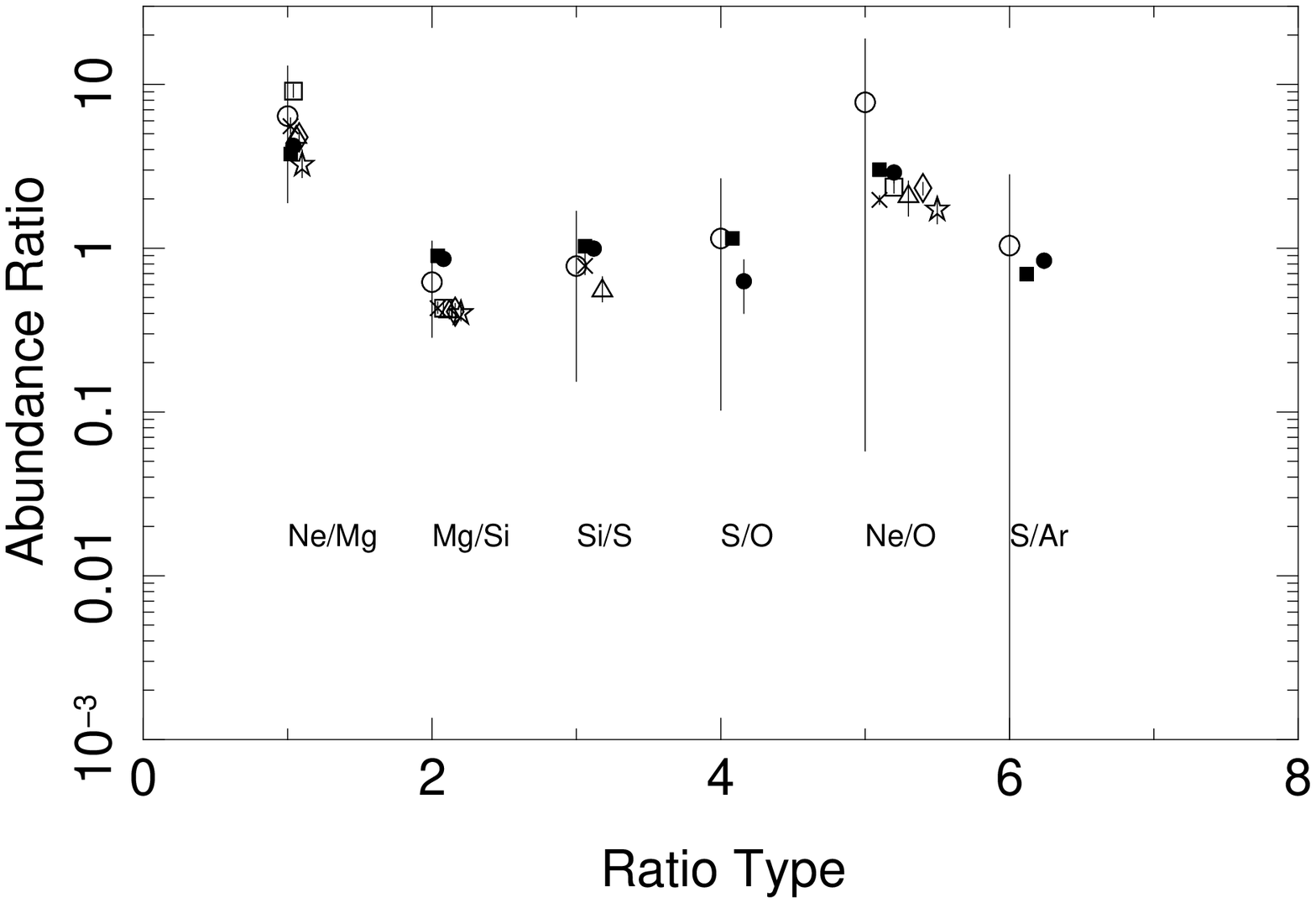}
\figcaption{The average abundance ratios for our Orion sources (circles) in comparison
with various M dwarfs with active coronae~\citep{liefke2008}. The symbols mean:
cross (YY Gem); square (AU Mic); trangle (EV Lac); diamond (AD Leo); star (Proxima
Cen); filled circle ($\sigma$ Gem); filled square (HR 1099).
\label{fig:mdwarfs}}
\smallskip

\section{Conclusions\label{conclusions}}

We observed some of the youngest CTTS in the ONC with ages between
0.1 and 10 Myr. All stars are very bright in X-rays with luminosities
significantly above 10$^{31}$ \ergsec and ratios of X-ray to bolometric 
luminosity of the order of -2.8. The X-ray surface flux increases 
significantly with young age even for stellar masses above 1.5 \Msun
together with average coronal loop sizes indicating an increase
in coronal activity during the CTTS phase. The column densities towards the
X-ray sources are lower than expected from measured extinction values
and lower than previously determined for the central massive star
$\theta^1$ Ori C, likely due to high local gas-to-dust ratios. 
The X-ray spectra are too hot to allow for significant R-ratio
determinations with current statistics, but on average the He-like
triplets are consistent with typical densities of coronal plasmas. Average
abundance ratios are also consistent with coronally active stars 
underlining the coronal nature of the observed X-ray emission.
The preliminary results of young coronally active PMS stars in this
analysis demonstrate the incredible potential of observing the ONC
stars with the \emph{Chandra} HETG to study details of the coronal evolution of CTTS.

\bibliographystyle{jwapjbib}

\input{ms_xar.bbl}
\end{document}

%% file: tab1.tex
\begin{center}
\caption{Line Fluxes of Major Lines used for Diagnostics\label{tab:linefits}}
\vskip 4pt
\begin{tabular}{lcccccc}
\hline
\hline
 Line & MT Ori  & LQ Ori & Par 1842 & V348 Ori & V229 Ori   & V1399 Ori \\ 
       &  (1)   &   (1)  &    (1)   &   (1)    &   (1)      &   (1)     \\ 
\hline 
 & & & & & & \\ 
Ar XVIII L$\alpha$ &  0.80 $^{  0.32 }_{  0.35 }$ &  0.34 $^{  0.20 }_{  0.23 }$ &  0.19 $^{  0.12 }_{  0.12 }$ &  0.04 $^{  0.16 }_{  0.21 }$ &  0.03 $^{  0.03 }_{  0.12 }$ &  0.19 $^{  0.17 }_{  0.18 }$ \\
Ar XVII r &  1.08 $^{  0.71 }_{  0.41 }$ &  0.28 $^{  0.19 }_{  0.22 }$ &  0.52 $^{  0.22 }_{  0.22 }$ &  0.22 $^{  0.02 }_{  0.27 }$ &  0.21 $^{  0.17 }_{  0.21 }$ &  0.31 $^{  0.20 }_{  0.23 }$ \\
S XVI L$\alpha$ &  0.75 $^{  0.35 }_{  0.40 }$ &  0.43 $^{  0.25 }_{  0.29 }$ &  0.37 $^{  0.30 }_{  0.30 }$ &  0.43 $^{  0.28 }_{  0.33 }$ &  1.48 $^{  0.33 }_{  0.37 }$ &  0.38 $^{  0.22 }_{  0.25 }$ \\
S XV r &  1.17 $^{  0.39 }_{  0.43 }$ &  0.58 $^{  0.31 }_{  0.39 }$ &  0.99 $^{  0.38 }_{  0.38 }$ &  0.48 $^{  0.34 }_{  0.43 }$ &  0.38 $^{  0.32 }_{  0.50 }$ &  0.57 $^{  0.25 }_{  0.29 }$ \\
S XV f &  0.70 $^{  0.35 }_{  0.40 }$ &  0.64 $^{  0.50 }_{  0.04 }$ &  0.45 $^{  0.26 }_{  0.26 }$ &  0.30 $^{  0.26 }_{  0.32 }$ &  0.58 $^{  0.26 }_{  0.30 }$ &  0.39 $^{  0.22 }_{  0.26 }$ \\
Si XIV L$\alpha$ &  1.60 $^{  0.27 }_{  0.29 }$ &  1.28 $^{  0.22 }_{  0.23 }$ &  0.98 $^{  0.21 }_{  0.21 }$ &  1.26 $^{  0.25 }_{  0.27 }$ &  1.14 $^{  0.21 }_{  0.22 }$ &  1.45 $^{  0.22 }_{  0.23 }$ \\
Si XIII r &  1.45 $^{  0.28 }_{  0.29 }$ &  0.65 $^{  0.19 }_{  0.21 }$ &  0.66 $^{  0.20 }_{  0.20 }$ &  0.72 $^{  0.28 }_{  0.34 }$ &  1.04 $^{  0.23 }_{  0.25 }$ &  1.03 $^{  0.20 }_{  0.22 }$ \\
Si XIII f &  0.95 $^{  0.24 }_{  0.25 }$ &  0.52 $^{  0.19 }_{  0.21 }$ &  0.32 $^{  0.17 }_{  0.17 }$ &  0.43 $^{  0.22 }_{  0.25 }$ &  0.42 $^{  0.15 }_{  0.17 }$ &  0.75 $^{  0.21 }_{  0.22 }$ \\
Mg XII L$\alpha$ &  1.87 $^{  0.28 }_{  0.30 }$ &  0.80 $^{  0.19 }_{  0.21 }$ &  0.47 $^{  0.18 }_{  0.18 }$ &  0.61 $^{  0.21 }_{  0.23 }$ &  0.92 $^{  0.20 }_{  0.22 }$ &  1.10 $^{  0.21 }_{  0.23 }$ \\
Mg XI r &  0.27 $^{  0.22 }_{  0.24 }$ &  0.32 $^{  0.19 }_{  0.21 }$ &  0.26 $^{  0.17 }_{  0.17 }$ &  0.77 $^{  0.25 }_{  0.28 }$ &  0.55 $^{  0.21 }_{  0.23 }$ &  0.71 $^{  0.20 }_{  0.22 }$ \\
Mg XI i &  0.76 $^{  0.28 }_{  0.30 }$ &  0.35 $^{  0.22 }_{  0.32 }$ &  0.03 $^{  0.00 }_{  0.00 }$ &  0.13 $^{  0.07 }_{  0.24 }$ &  0.26 $^{  0.19 }_{  0.21 }$ &  0.42 $^{  0.24 }_{  0.21 }$ \\
Mg XI f &  0.20 $^{  0.20 }_{  0.25 }$ &  0.00 $^{  0.00 }_{  0.00 }$ &  0.33 $^{  0.00 }_{  0.00 }$ &  0.29 $^{  0.20 }_{  0.24 }$ &  0.27 $^{  0.18 }_{  0.21 }$ &  0.47 $^{  0.18 }_{  0.20 }$ \\
Ne X L$\alpha$ & 24.12 $^{  4.03 }_{  1.87 }$ & 20.06 $^{  2.73 }_{  3.40 }$ &  8.86 $^{  1.04 }_{  1.04 }$ & 11.59 $^{  4.48 }_{  1.38 }$ & 12.64 $^{  2.21 }_{  1.29 }$ &  8.33 $^{  1.11 }_{  1.17 }$ \\
Ne IX r &  4.33 $^{  2.16 }_{  1.14 }$ &  4.20 $^{  4.00 }_{  2.06 }$ &  2.64 $^{  0.81 }_{  0.81 }$ &  2.27 $^{  0.98 }_{  1.15 }$ &  2.40 $^{  1.20 }_{  0.95 }$ &  1.12 $^{  0.60 }_{  0.73 }$ \\
Ne IX i &  0.83 $^{  0.57 }_{  0.69 }$ &  0.84 $^{  0.57 }_{  0.70 }$ &  0.85 $^{  0.54 }_{  0.54 }$ &  1.52 $^{  0.74 }_{  0.92 }$ &  1.07 $^{  0.56 }_{  0.71 }$ &  0.30 $^{  0.10 }_{  0.54 }$ \\
Ne IX f &  2.42 $^{  0.83 }_{  0.97 }$ &  1.73 $^{  0.74 }_{  0.88 }$ &  1.74 $^{  0.72 }_{  0.72 }$ &  1.11 $^{  0.68 }_{  0.88 }$ &  2.07 $^{  0.78 }_{  0.94 }$ &  1.03 $^{  0.55 }_{  0.75 }$ \\
O VIII L$\alpha$ & 16.28 $^{  4.41 }_{  5.25 }$ & 29.59 $^{  5.86 }_{  6.64 }$ &  5.21 $^{  2.46 }_{  2.46 }$ &  0.20 $^{  0.00 }_{  0.00 }$ &  2.38 $^{  1.72 }_{  2.74 }$ &  8.28 $^{  2.87 }_{  3.56 }$ \\
O VII r &  6.01 $^{  4.52 }_{  7.34 }$ &  0.20 $^{  0.00 }_{  3.50 }$ &  7.65 $^{  4.73 }_{  4.73 }$ &  2.55 $^{  2.35 }_{  6.92 }$ &  0.00 $^{  0.20 }_{  2.35 }$ &  4.94 $^{  3.79 }_{  6.07 }$ \\
O VII i &  3.01 $^{  2.96 }_{  5.75 }$ &  0.20 $^{  0.00 }_{  2.67 }$ &  4.14 $^{  3.46 }_{  3.46 }$ &  6.35 $^{  5.03 }_{  8.69 }$ & 12.81 $^{  7.19 }_{ 10.60 }$ &  2.37 $^{  2.17 }_{  5.30 }$ \\
O VII f &  7.17 $^{  5.27 }_{  8.65 }$ &  3.70 $^{  3.64 }_{  7.37 }$ &  5.01 $^{  3.99 }_{  3.99 }$ &  3.54 $^{  3.54 }_{  7.11 }$ &  0.20 $^{  0.00 }_{  2.90 }$ & 13.19 $^{  7.18 }_{ 10.36 }$ \\
N VII L$\alpha$ &  6.42 $^{  4.38 }_{  7.01 }$ &  4.99 $^{  3.63 }_{  5.82 }$ &  0.00 $^{  0.00 }_{  0.00 }$ &  0.00 $^{  0.00 }_{  2.34 }$ &  6.17 $^{  5.33 }_{  8.05 }$ &  8.36 $^{  5.21 }_{  7.62 }$ \\
& & & & &  \\ 
\hline 
\end{tabular} 
\end{center} 
(1) 10$^{-6}$ photons cm$^{-2}$ s$^{-1}$ 